\newcommand\Mtens{\mathbf{M}(\boldsymbol{x})}
\newcommand\ximg{\textbf{\emph{x}}}
\newcommand\usrc{\textbf{\emph{u}}}

\documentclass[apj,numberedappendix]{emulateapj}
\usepackage{amsmath}
\usepackage{multirow}
\usepackage{natbib}
\shorttitle{OPTIMAL CONFIGURATIONS FOR LENSING HIGH-Z GALAXIES}
\shortauthors{WONG ET AL.}

\begin{document}
\title{OPTIMAL MASS CONFIGURATIONS FOR LENSING HIGH-REDSHIFT GALAXIES}
\author{
Kenneth C. Wong\altaffilmark{1}, 
S. Mark Ammons\altaffilmark{2},
Charles R. Keeton\altaffilmark{3}, 
and Ann I. Zabludoff\altaffilmark{1}}
\altaffiltext{1}{Steward Observatory, University of Arizona, 933 North Cherry Avenue, Tucson, AZ 85721}
\altaffiltext{2}{Lawrence Livermore National Laboratory, 7000 East Avenue, Livermore, CA 94550}
\altaffiltext{3}{Department of Physics and Astronomy, Rutgers University, 136 Frelinghuysen Road, Piscataway, NJ 08854}

\begin{abstract}
We investigate the gravitational lensing properties of lines of sight containing multiple cluster-scale halos, motivated by their ability to lens very high-redshift ($z \sim 10$) sources into detectability. We control for the total mass along the line of sight, isolating the effects of distributing the mass among multiple halos and of varying the physical properties of the halos.  Our results show that multiple-halo lines of sight can increase the magnified source-plane region compared to the single cluster lenses typically targeted for lensing studies, and thus are generally better fields for detecting very high-redshift sources.  The configurations that result in optimal lensing cross sections benefit from interactions between the lens potentials of the halos when they overlap somewhat on the sky, creating regions of high magnification in the source plane not present when the halos are considered individually.  The effect of these interactions on the lensing cross section can even be comparable to changing the total mass of the lens from $10^{15} M_{\odot}$ to $3\times10^{15} M_{\odot}$.  The gain in lensing cross section increases as the mass is split into more halos, provided that the lens potentials are projected close enough to interact with each other.  A nonzero projected halo angular separation, equal halo mass ratio, and high projected halo concentration are the best mass configurations, whereas projected halo ellipticity, halo triaxiality, and the relative orientations of the halos are less important.  Such high mass, multiple-halo lines of sight exist in the SDSS.
\end{abstract}


\section{INTRODUCTION} \label{sec:intro}
Detecting very high-redshift galaxies and determining their abundance and star-formation properties is one of the foremost challenges in observational cosmology today.  These early galaxies contain the first generation of stars in the universe, and their UV emission is thought to contribute to the reionization of neutral hydrogen during this epoch.  However, observing these faint objects is extremely difficult.  Past observational studies of $z \gtrsim 7$ galaxies in blank fields using the Lyman-break dropout technique \citep[e.g.,][]{bouwens2004,bouwens2008,bouwens2010a,bouwens2010b,bouwens2011a,bouwens2011b,labbe2006,labbe2010a,labbe2010b,henry2007,oesch2009,oesch2010b,oesch2012a,oesch2012b,bunker2010,castellano2010,ouchi2009,wilkins2010,wilkins2011a,wilkins2011b,capak2011,lorenzoni2011,trenti2011,trenti2012,yan2011a,yan2011b,hsieh2012}, narrow-band searches for Lyman-$\alpha$ emission \citep[e.g.,][]{ota2008,sobral2009,hibon2010,hibon2012,tilvi2010,krug2012}, or full SED fitting to broadband photometry \citep[e.g.,][]{finkelstein2010,finkelstein2011,mclure2010,mclure2011} have required a large investment of observing time on {\it HST} and/or large ($\gtrsim 4$-m) ground-based telescopes and have detected only a handful of candidates.

Gravitational lensing by galaxy clusters aids in the detection of these sources by magnifying the background source population.  A number of surveys toward known lensing clusters have revealed a few high-redshift galaxy candidates \citep[e.g.,][]{kneib2004,pello2004,schaerer2005,richard2006,richard2008,stark2007,bradley2008,bradley2012,zheng2009,laporte2011,hall2012}.  While lensing makes background sources easier to detect by mapping regions from the source plane to a larger area in the lens plane, there is a corresponding loss of volume probed in the source plane.  These two effects (gain in depth, loss in volume) affect the efficiency of searches for high-redshift galaxies in different ways.  Addressing the effects of lensing magnification on the detectability of high-redshift sources \citep[e.g.,][]{bouwens2009,maizy2010} is subject to uncertainties in the source properties, such as the poorly-constrained shape of the $z \geq 7$ galaxy luminosity function (LF), the lens properties, and the observational strategy.  The efficiency of massive cluster-scale halos at magnifying a background source population is affected by halo mass, ellipticity, orientation, and substructure, as well as by uncorrelated structures along the line of sight \citep[LOS; ][]{wambsganss2005,hilbert2007,puchwein2009}.

The lensing properties of a given LOS, or ``beam'', depend not only on the total integrated mass along the beam, but also on the distribution of that mass.  Ideally, that mass should be distributed in a way that maximizes the \'{e}tendue, the area in the source plane over which significant magnification occurs.  Selection of cluster-scale lenses through X-ray observations and weak lensing studies, among other methods, has often yielded fields where a single massive structure dominates the beam, while any smaller LOS structures are ignored.  Yet, beams could exist where {\it multiple} cluster-scale structures maximize the contribution to the lens potential, enhancing the detectability of lensed background sources over that of a single lensing cluster, even if the total integrated mass in the beam is the same.

A few studies have explored the theoretical lensing properties of lines of sight containing multiple clusters \citep{crawford1986,bertin2001}, often focusing on cluster-cluster strong lensing \citep[e.g.][]{cooray1999}, but few such configurations have been observed and characterized in terms of the properties that maximize their lensing strength.  Possible chance alignments in cluster observations were also suggested by \citet{molinari1996} and \citet{wang1997}.  Serendipitous discoveries have since shown that chance alignments of clusters exist in the observable universe \citep{blakeslee2001,athreya2002,zitrin2012}.  However, the lensing properties of such multi-cluster beams relative to single clusters have not been well-studied, and the likelihood of these alignments is not well-known.  Distributing the mass in a beam among multiple projected lensing halos may lead to beneficial interactions of the lens potentials and thus higher \'{e}tendue.  If these beams are better at magnifying distant sources than typical lensing clusters, it is important to identify the characteristics that make them powerful lenses and to quantify their frequency in the universe.

The physical properties of the halos in multi-halo beams can also impact their ability to lens high-redshift sources.  Several past studies have examined the effects of halo concentration, shape, and substructure, although in the context of lensing $z \sim 1-2$ galaxies into arcs.  \citet{dalal2004} and \citet{mandelbaum2009} find a strong correlation between halo mass and the lensing cross section, consistent with arcs in galaxy clusters and not in lower mass systems.  As a result of numerical studies of halo concentration, ellipticity, and triaxiality, several groups \citep[e.g.,][]{bartelmann1995,meneghetti2003,oguri2003} find that the lensing cross section increases as the mass distribution of the halo becomes more centrally concentrated (either through a steepening of the inner density slope or an increase in the projected halo concentration).  These studies have focused on single halos and generally do not seek to identify the properties of the very best and realistic lensing configurations for very high-redshift galaxies.

In this paper, we perform a Monte Carlo analysis to model beams containing multiple cluster-scale halos and evaluate their lensing properties in comparison with single cluster lenses of equivalent total mass.  By controlling for total mass, we isolate the gain due purely to the mass configuration $-$ the number of halos and their physical properties.  We apply physically-motivated priors to ensure that our beams are realistic.

We describe the details of our lensing analysis and outline our methodology for characterizing the best lensing beams in \S~\ref{sec:beams}.  In \S~\ref{sec:toymodel}, we describe the models we use to represent various mass configurations and test their lensing properties.  We present our results in \S~\ref{sec:results}.  We discuss the interpretation of our results in \S~\ref{sec:discussion}, as well as possible caveats and sources of uncertainty.  In addition, we discuss the implications of our findings on the detection of high-redshift sources and the likelihood of finding such mass configurations in the universe.  Our main conclusions are summarized in \S~\ref{sec:conclusions}.  Throughout this paper, we assume a $\Lambda$CDM cosmology with $\Omega_{m} = 0.274$, $\Omega_{b} = 0.045$, $\Omega_{\Lambda} = 0.726$, and $H_{0} = 71$ km s$^{-1}$ Mpc$^{-1}$.

\section{CHARACTERIZING THE BEST MASS CONFIGURATIONS FOR LENSING} \label{sec:beams}
The lensing properties of a particular mass configuration for a high-redshift source plane can be influenced by several physical characteristics.  The total mass along the line of sight clearly is important, as well as the manner in which that mass is distributed, both in projection and redshift.  In addition, the structural properties of the halos containing the mass will have an effect on the lensing.  In this section, we describe our lensing calculations and our metric for evaluating the lensing properties of our simulated beams, as well as introduce our tests for looking at the effects of the various physical properties of their mass configurations.

\subsection{Lensing Analysis} \label{subsec:lensing}
All of our lensing calculations are performed using an updated version of the software described in \citet{keeton2001a}.  Here, we describe methodology for determining the lensing properties of a given mass configuration.  We select a source redshift of $z_{s} = 10$, which is close to the current limits of what is currently feasible with existing observatories \citep{bouwens2011a}.

To study the lensing properties of various mass configurations on a $z  =10$ source plane, we construct a uniform 5\arcmin$\times$5\arcmin~grid of small circular sources at 1\arcsec~intervals on the source plane.  In principle, we could use elliptical sources \citep{keeton2001b}, but this adds an additional complication, so we assume circular sources.  We then trace these sources through the mass configuration.  The result is a distribution of image positions that will always contain a number of points greater than or equal to the original source plane grid due to multiple imaging of sources within the lensing caustics.  For each source, we tabulate the images associated with it, as well as the elements of the magnification tensor, $\Mtens$, for each image.  The total image magnification at \ximg~is just the determinant of $\Mtens$.

The axis ratio of the image (assuming an intrinsically circular source) can be approximated as the ratio of the eigenvalues of the magnification tensor \citep[e.g.,][]{schneider1992,fedeli2006}, $a/b \approx \lambda_{1}/\lambda_{2}$, where $a$ and $b$ are the major and minor axes of the lensed image and $\lambda_{1}$ and $\lambda_{2}$ are the maximum and minimum eigenvalues of $\Mtens$, respectively.  This approximation should be valid for small circular sources, but can break down for extended sources when the magnification tensor changes appreciably over the area of the source \citep{fedeli2006}.

The magnification tensor at each image position is calculated numerically, accounting for the effects of multiple lensing planes.  We provide a brief discussion of the lensing formalism behind these calculations since we are often dealing with multiple halos at different redshifts (see also \citet{kochanek1988} and McCully et al. in preparation).  Angular positions in the source plane and image plane are denoted \usrc~and \ximg, respectively.

For simplicity, first consider the case of two halos at the same redshift so they can be treated as being in the same lens plane.  The lens potential is the superposition of the potentials of the two halos,
\begin{equation} \label{eq:lens}
\phi_{\rm tot}(\ximg) = \phi_1(\ximg) + \phi_2(\ximg).
\end{equation}
The lens equation is 
\begin{equation} \label{eq:lensA}
\begin{split}
\usrc &= \ximg - \nabla\phi_{\rm tot} \\
&= \ximg - {\boldsymbol \alpha}_1(\ximg) - {\boldsymbol \alpha}_2(\ximg),
\end{split}
\end{equation}
where ${\boldsymbol \alpha}_1=\nabla\phi_1$ and ${\boldsymbol \alpha}_2=\nabla\phi_2$ are the deflections for the two halos.  The inverse magnification tensor is
{\setlength{\arraycolsep}{0pt}
\begin{equation} \label{eq:mtensA}
\Mtens^{-1} = \left[\begin{array}{cc}
    1 - \kappa_1 - \kappa_2 - \gamma_{c1} - \gamma_{c2} & - \gamma_{s1} - \gamma_{s2} \\
    - \gamma_{s1} - \gamma_{s1} & 1 - \kappa_1 - \kappa_2 + \gamma_{c1} + \gamma_{c2}
  \end{array}\right],
\end{equation}
}
where $\kappa$ indicates the convergence (the isotropic magnification term) while $\gamma_c$ and $\gamma_s$ indicate the two components of shear (the anisotropic magnification term).  In general, these three quantities will have similar amplitudes $-$ not necessarily identical, but within a factor of a few of each other.  The magnification at a given image position is then
\begin{equation} \label{eq:mag}
\begin{split}
\mu &= \mathrm{det}(\Mtens) = \frac{1}{\mathrm{det}(\Mtens^{-1})} \\
&= \left[ (1-\kappa_1-\kappa_2)^2 - (\gamma_{c1}+\gamma_{c2})^2 - (\gamma_{s1}+\gamma_{s2})^2 \right]^{-1}.
\end{split}
\end{equation}
The convergence and shear terms from both halos enter the equation at the same lowest order.

Consider a scenario where the two halos are separated enough that their caustics are distinct.  We would like to examine how the lens potentials of the halos interact with each other and affect the magnification in regions near one of them.  Suppose we are reasonably close to halo \#1.   Equation~\ref{eq:mag} indicates that the change in the magnification across the field due to the presence of halo \#2 can be due to both its density profile and shear effects.  If halo \#2 is far enough away that its convergence and shear are relatively small, we can do a Taylor series expansion in the quantities from halo \#2,
\begin{equation} \label{eq:muser}
\mu \approx \mu_1 + 2 \mu_1^2 \left[(1-\kappa_1)\kappa_2 + \gamma_{c1} \gamma_{c2}
    + \gamma_{s1} \gamma_{s2}\right] + \ldots,
\end{equation}
where $\mu_1 = [(1-\kappa_1)^2-\gamma_{c1}^2-\gamma_{s1}^2]^{-1}$ is the magnification from halo \#1 alone.  The extra term in Equation~(\ref{eq:muser}) does {\it not} have the form of the magnification expected from halo \#2 alone.  This is the statement that $-$ unlike the lens potentials, deflections, convergence terms, and shear terms $-$ magnifications do not simply add linearly because magnification is a nonlinear function of the potential.  This term is not necessarily positive since $\gamma_{c}$ and $\gamma_{s}$ can have either sign, although in practice the presence of halo \#2 generally increases the magnification rather than decreases it.

Now consider the case of two halos at different redshifts.  Let halo \#1 be in the foreground and halo \#2 be in the background.  Now the lens equation has the form
\begin{equation} \label{eq:lensB}
\usrc = \ximg - {\boldsymbol \alpha}_1(\ximg) - {\boldsymbol \alpha}_2(\ximg-\beta_{12}{\boldsymbol \alpha}_1(\ximg)),
\end{equation}
where $\beta_{12}$ is a ratio of angular diameter distances between combinations of the observer, the two lens planes, and the source plane:
\begin{equation} \label{eq:beta}
\beta_{12} = \frac{D_{12} D_s}{D_2 D_{1s}}.
\end{equation}
Note in Equation~\ref{eq:lensB} that ${\boldsymbol \alpha}_1$ appears {\it inside the argument} of ${\boldsymbol \alpha}_2$.  This occurs because the location where a light ray intersects plane \#2 is shifted from the position of the light ray in the image plane due to the presence of the foreground lens.  The inverse magnification tensor is
{\setlength{\arraycolsep}{0pt}
\begin{equation} \label{eq:mtensB}
\begin{split}
\Mtens&^{-1} = 
  \left[\begin{array}{cc}
    1 - (\kappa_1+\gamma_{c1}) & - \gamma_{s1} \\
    - \gamma_{s1} & 1 - (\kappa_1-\gamma_{c1})
  \end{array}\right] \\
  & - 
  \left[\begin{array}{cc}
    \kappa_2 + \gamma_{c2} & \gamma_{s2} \\
    \gamma_{s2} & \kappa_2 - \gamma_{c2}
  \end{array}\right]
  \left[\begin{array}{cc}
    1 - \beta_{12}(\kappa_1+\gamma_{c1}) & - \beta_{12} \gamma_{s1} \\
    - \beta_{12} \gamma_{s1} & 1 - \beta_{12} (\kappa_1-\gamma_{c1})
  \end{array}\right]
\end{split}
\end{equation}
}
This is actually even more complicated than it appears because (like ${\boldsymbol \alpha}_2$) the functions $\kappa_2$, $\gamma_{c2}$, and $\gamma_{s2}$ are all evaluated at the position $\ximg-\beta_{12}{\boldsymbol \alpha}_1(\ximg)$.  The expression for $\mu$ is similarly complicated.

The formalism can be extended to an arbitrary number of lens planes \citep{schneider1992,petters2001}.  The inverse magnification tensor for the $j$th lens plane of a given line of sight can be calculated through the recursion relation
\begin{equation} \label{eq:recursion}
\Mtens_{j}^{-1} = {\bf I} - \sum_{i=1}^{j-1} \beta_{ij} {\boldsymbol \Gamma}_{i} \Mtens_{i}^{-1},
\end{equation}
where {\bf I} is the $2\times2$ identity matrix, $\Mtens_{1}^{-1} = {\bf I}$, and ${\boldsymbol \Gamma}_{i}$ is the ``shear tensor'' for plane $i$,
\begin{equation} \label{eq:sheartens}
  {\boldsymbol \Gamma}_i = \left[\begin{array}{cc}
    \kappa_i + \gamma_{ci} & \gamma_{si} \\
    \gamma_{si} & \kappa_i - \gamma_{ci}
  \end{array}\right].
\end{equation}
The inverse magnification tensor for the entire mass distribution is given by Equation~\ref{eq:recursion} evaluated at the source plane, $\Mtens_{s}^{-1}$, and the image magnification and axis ratio can then be evaluated.  We will revisit this formalism when interpreting our results.

\subsection{Quantifying the Lensing Properties of Different Mass Configurations} \label{subsec:quantify}
There are a number of quantities we can use to determine how good a particular line of sight is for lensing.  The relative merits of each quantity depends on our ultimate objective.  For the purposes of this work, we are primarily interested in finding the lines of sight that are most effective at magnifying very high-redshift sources ($z \sim 10$) into detectability.

A simple and useful metric for evaluating the lensing power of a particular mass configuration is the cross section for magnifying a source above some minimum threshold value.  This cross section, which we denote $\sigma_{\mu}$, is the area in the source plane over which a lensed source's brightest image is magnified above the chosen threshold magnification.  $\sigma_{\mu}$ is a convenient quantity because it makes no assumptions about the properties of the source population (aside from its redshift) or the observational setup, but is dependent only on the mass configuration.  We choose to define $\sigma_{\mu}$ in terms of source plane area rather than image plane area because the former is a better metric for evaluating the quality of a particular line of sight in terms of maximizing the potential number of high-redshift detections.  Although area in the image plane with high $\mu$ may be a relevant quantity, its usefulness depends on the specific details of the observations designed to study these lines of sight due to source size effects, and is therefore less general.

We define $\sigma_{\mu}$ in terms of the magnification of the brightest source image rather than total magnification because it is the detectability of individual images that interests us.  If a source is multiply imaged such that it has a high total magnification but no individual image is magnified enough to be detectable, it is of no use.  Throughout this paper, we define $\sigma_{\mu}$ as the source plane area where a source's brightest image has a magnification $\mu \geq 3$.  This value is somewhat arbitrary, although there are indications that the number of high-redshift detections in lensing fields, while relatively flat as a function of magnification, broadly peaks in regions with $\mu \sim 3$ (Ammons et al. in preparation).  We ultimately find that our general conclusions are not affected by our choice of magnification threshold, and we discuss the implications of changing this threshold in Appendix~\ref{app:muthreshold}.  Formally, $\sigma_{\mu}$ is defined as
\begin{equation} \label{eq:sigmamu}
\sigma_{\mu} = \int_{\Omega} \mathcal{H} \left( \rm{max}(\mu) - \mu_{\emph{t}} \right) d\Omega,
\end{equation}
where $\Omega$ is area in the source plane, $\mathcal{H}$ is the Heaviside step function, max($\mu$) is the highest magnification of any individual image associated with a given source position, and $\mu_{t}$ is the threshold magnification, which we choose to be $\mu_{t} = 3$.  $\sigma_{\mu}$ can be thought of as \'{e}tendue $-$ the area in the source plane that would get magnified above a certain value of interest.

\begin{figure}
\centering
\plotone{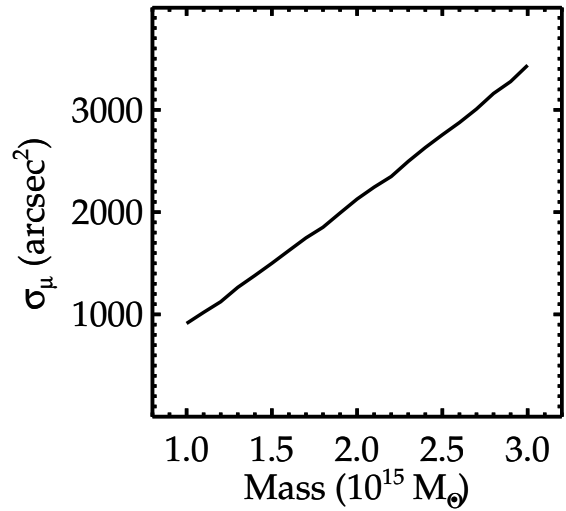}
\caption{The lensing cross section, $\sigma_{\mu}$, as a function of mass for a spherical halo at $z = 0.5$.  The cross section increases monotonically with mass, reflecting the fact that more massive halos are better lenses.  \label{fig:sigmamass}}
\end{figure}

Figure~\ref{fig:sigmamass} shows $\sigma_{\mu}$ as a function of mass for a single spherical NFW halo \citep{navarro1996} at $z = 0.5$, with structural properties at each mass given by the median relations of \citet{zhao2009}.  The halo mass is defined as $M_{200}$, the mass enclosed within a sphere whose mean density is 200 times the mean matter density at that redshift.  The mass range is chosen to represent realistic sizes for massive cluster lenses.  As expected, the cross section increases monotonically with mass, changing by $\gtrsim$3$\times$ for a corresponding factor of 3 change in mass over this range.  This reflects the fact that more massive halos are generally better lenses.  This figure also gives a sense of the value of $\sigma_{\mu}$ expected for massive clusters and how it changes with the addition of more mass.  We later use this model to roughly compare gains in $\sigma_{\mu}$ arising from the mass configuration alone to those from the addition of more total mass.

\section{LINE-OF-SIGHT MASS MODELS} \label{sec:toymodel}
The halos used in our analysis are parameterized as NFW profiles, an assumption that facilitates comparison with past and on-going work on massive cluster lenses and is supported by several observational studies \citep[e.g.,][]{kneib2003,katgert2004,comerford2006,shu2008}.  We neglect the effects of substructure and baryonic matter in our calculations, although we comment on this in \S~\ref{subsec:uncertainties}.

\subsection{Priors on Halo Properties} \label{subsec:halos}
We generate halos with redshifts drawn from a distribution in the range $0.1 \leq z \leq 0.9$.  We choose this range because we expect the majority of massive cluster-scale lenses to be at these low-to-intermediate redshifts due to the evolution of the halo mass function.  At very low redshift ($z \lesssim 0.1$), there is little volume and the surface density of cluster halos becomes small (assuming a shallower-than-isothermal mass profile), making these halos less effective lenses.  At higher redshifts ($z \gtrsim 1$), the lensing geometry works against gains in magnification because the angular diameter distance ratios among the observer, lens, and source become unfavorable for lensing high-redshift sources.  Furthermore, cluster-scale halos are more rare and the evolution of cluster properties has not been well constrained by observations at these redshifts.  Our redshift range encompasses the theoretical peak in the lens redshift distribution for clusters lensing giant arcs \citep[$z \sim 0.3-0.4$;][]{bartelmann1998}, as well as the observed peak in the redshifts of arc-producing lensing clusters from the RCS2 dataset \citep[$z \gtrsim 0.6$;][]{gladders2003,hennawi2007}.  The redshift distribution from which we draw is weighted by the number of halos with masses greater than $10^{14} M_{\odot}$ expected within a comoving volume element defined by the solid angle of our fiducial aperture of 3\farcm5 (see below).  This procedure ends up populating the full redshift range ($0.1 \leq z \leq 0.9$), although not uniformly.  Our procedure includes the possibility that multiple halos lie at the same redshift, as would subclusters in clusters of galaxies, although we do not account for physical interactions among the halos as might occur in subcluster mergers.

The number density of halos at a given mass and redshift is determined using the publicly-available halo mass function code of \citet{tinker2008}, which assumes the \citet{eisenstein1998} matter transfer function as the default.  In this calculation, we also assume our fiducial cosmological parameters, along with $\sigma_{8} = 0.8$, $\delta_{crit} = 1.686$, and a primordial power spectrum index of 0.963.  We define halos to have an overdensity of $\Delta = 200$ relative to the mean background density at that redshift.  We assign each halo a mass drawn from the mass function at its given redshift, requiring that the halo has a mass of at least $10^{14} M_{\odot}$, roughly the mass of a Virgo-scale galaxy cluster.

We perform a Monte Carlo analysis to generate lines of sight that contain $N$ halos, where $N$ is the number of distinct halos that we are testing in that particular set of trials.  We examine a control sample of beams containing $N = 1$ halo (\S~\ref{subsec:1halo}), then explore beams with multiple halos for comparison (\S~\ref{subsec:2halo} and \S~\ref{subsec:3+halo}).  For each trial, we randomly draw redshifts for $N$ halos using the procedure described above.  We then randomly assign masses to these halos using the mass function at each halo redshift until we find a realization where the total mass across all $N$ halos is within 1\% of some fiducial beam mass.  We choose this fiducial mass to be $2 \times 10^{15} M_{\odot}$ which is roughly the virial mass of massive clusters like Coma \citep{kubo2007} and Abell 1689 \citep{umetsu2008}.

The angular positions of the halos in a particular realization are drawn randomly to approximate the distribution of physically uncorrelated halos in such a line of sight within a fixed aperture.  We choose an aperture size of 3\farcm5 in diameter, which is a separation large enough that halos with our chosen range of properties generally behave as independent lenses and do not gain from interactions with other halos (see \S~\ref{subsec:multihalo_results}).  We place one halo at the origin and randomly distribute the remaining $N-1$ halos uniformly across a disk of radius 3\farcm5.  We then calculate the centroid and diameter of the minimum enclosing circle of the distribution.  If the diameter is less than 3\farcm5, we retain that configuration, otherwise we repeat the procedure.  After finding a suitable configuration for a particular trial, the field across which we perform our lensing analysis is re-centered on the centroid of the minimum enclosing circle.

For beams containing multiple halos, this procedure assumes that the positions of cluster-scale halos on the sky are uncorrelated on scales smaller than a few arcminutes.  In detail, there may be correlations between halos at small redshift separation due to large scale structure (e.g. filaments, superclusters).  However, accounting for this would require detailed knowledge of the evolution of the cluster correlation function on $\sim$arcminute scales with redshift, which is not well-constrained.  Such correlations will be investigated in a future work using cosmological simulations (French et al. in preparation), but we ignore this effect for our analysis.

To determine the relevant physical properties of the halos for our lensing calculations, we use the publicly-available halo evolution code based on the models of \citet{zhao2009}.  Using this code, we determine the median virial radius $r_{vir}$, concentration parameter $c_{vir}$, scale radius $r_{s}$, and characteristic density $\rho_{s}$ as a function of halo mass and redshift and apply these to our halos, assuming the same cosmological parameters as described above.  We note that recent results suggest a possible reversal in the halo mass-concentration relation at the cluster scale \citep[e.g.,][]{klypin2011,prada2011}, which we do not account for.  However, such a trend would affect both our control sample and multi-halo beams in similar ways, so the relative properties of these lines of sight should not be significantly affected.  We include an intrinsic scatter in $c_{vir}$ of $\sigma_{\log{c}} = 0.14$ \citep{bullock2001}, which affects the values of $r_{s}$ and $\kappa_{s}$ (the characteristic surface density) that go into the lensing calculation but keeps virial mass and radius fixed.

Lensing clusters have been found to have an orientation bias such that the major axis of lensing-selected halos tends to be preferentially aligned with the line of sight from the observer.  This has the effect of increasing their projected concentration in the plane of the sky \citep{corless2007,hennawi2007,broadhurst2008,oguri2009}.  We are not looking specifically at lines of sight that have been predetermined to be good lenses, but triaxiality and projection effects can certainly affect the lensing properties of these mass configurations.  For this reason, we assume that our halos are triaxial, with axis ratios drawn from Gaussian distributions determined from simulations performed by \citet{shaw2006}.  We use the results from their simulations that include mass structures just beyond the virial radii of their halos, which give axis ratios of $b/a = 0.79 \pm 0.121$ and $c/a = 0.678 \pm 0.113$, where $a$, $b$, and $c$ are the major, intermediate, and minor semi-axes of the halo.  We then give our halos a random orientation in a spherical coordinate system and project the mass distributions onto the plane of the sky to get their projected ellipticities and position angles.  We can also determine the effects of this projection on the relevant physical quantities for lensing (see Appendix~\ref{app:projection} for details).  We note that our assumed NFW halos have elliptically symmetric density profiles (as opposed to elliptically symmetric lens potentials) for which the potentials do not have analytic solutions and are instead computed numerically.

Any values of lensing quantities given here are specific to the mass models with the properties used in our analysis.  In general, the diversity of possible mass configurations in the universe likely results in a large range of possible lensing properties and implied effects on the detection of high-redshift sources.  We have imposed several physically-motivated priors to reflect realistic mass distributions, but we caution against using the absolute numbers derived in this analysis, instead focusing on comparing the relative values of the lensing properties of the multiple-halo lensing beams to that of the single-halo case.

\subsection{Single Halo Control Sample} \label{subsec:1halo}
\begin{figure*}
\centering
\plotone{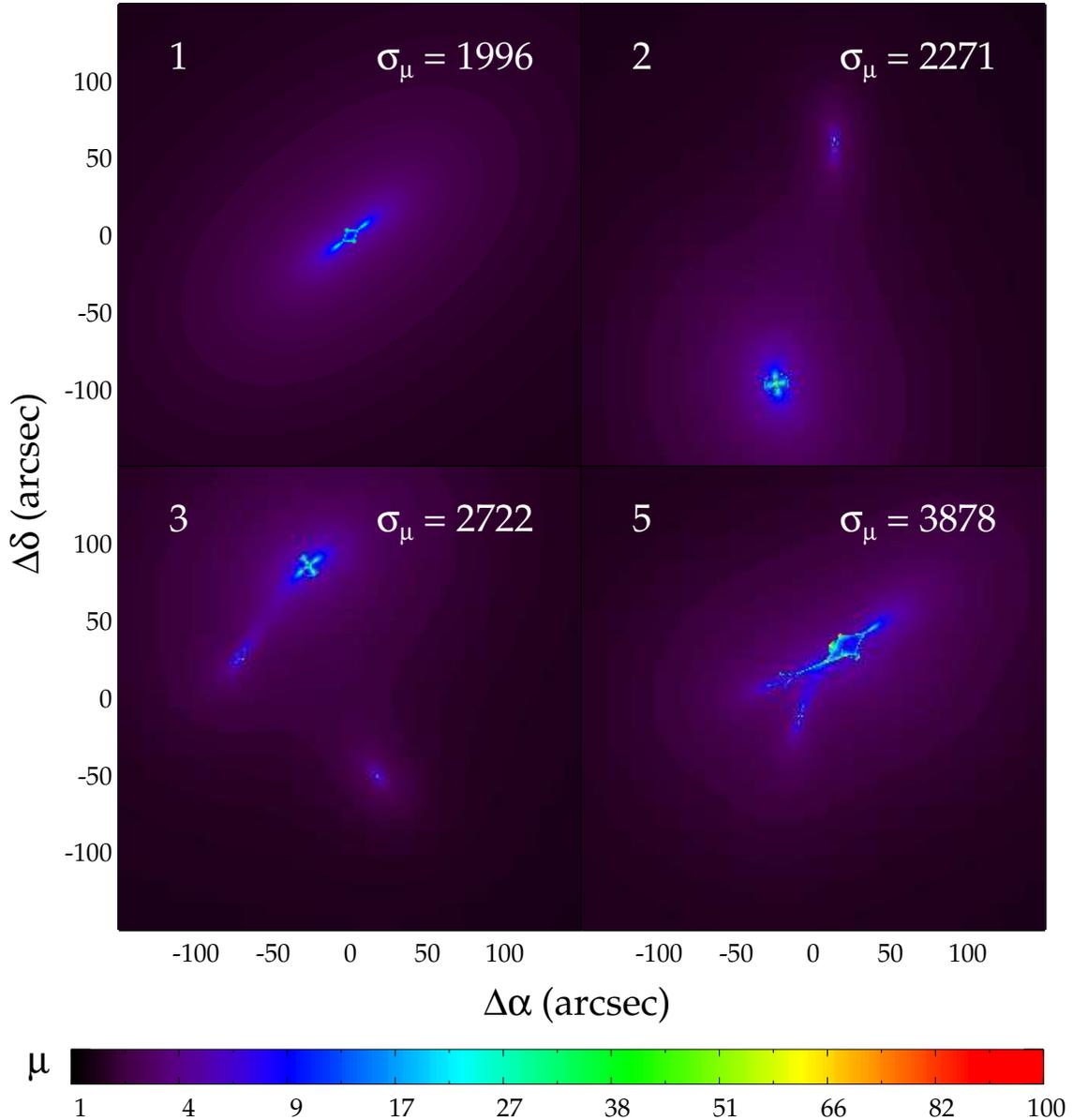}
\caption{Sample $z = 10$ source plane magnification maps over a $300\arcsec \times 300\arcsec$ field for lines of sight with close to median $\sigma_{\mu}$ values containing a single halo (top left), two halos (top right), three halos (bottom left), and five halos (bottom right).  The colors correspond to the magnification of the single most magnified image of a small source at each position on the map.  The color map is logarithmic in the magnification.  The values of $\sigma_{\mu}$ (arcsec$^{2}$) for these particular configurations are indicated in the top right corner of each panel.  These median configurations show that increasing the number of halos results in an increase in $\sigma_{\mu}$. \label{fig:magmap}}
\end{figure*}

\begin{figure*}
\centering
\plotone{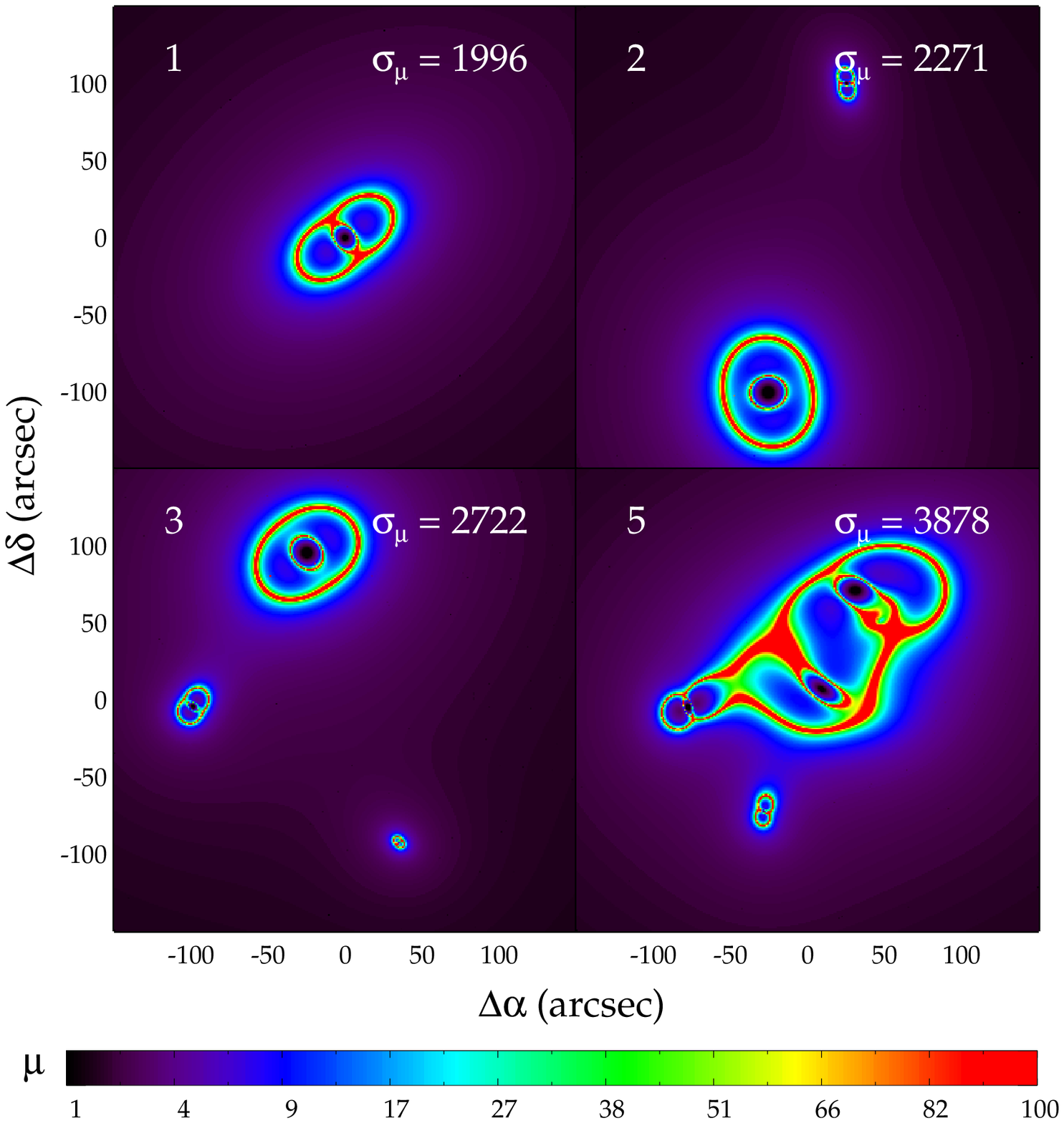}
\caption{Same as Figure~\ref{fig:magmap} but the maps in the image plane are now shown.  The magnification maps shown here correspond to the same mass distributions as in Figure~\ref{fig:magmap}.  \label{fig:magmapimg}}
\end{figure*}

\begin{figure*}
\centering
\plotone{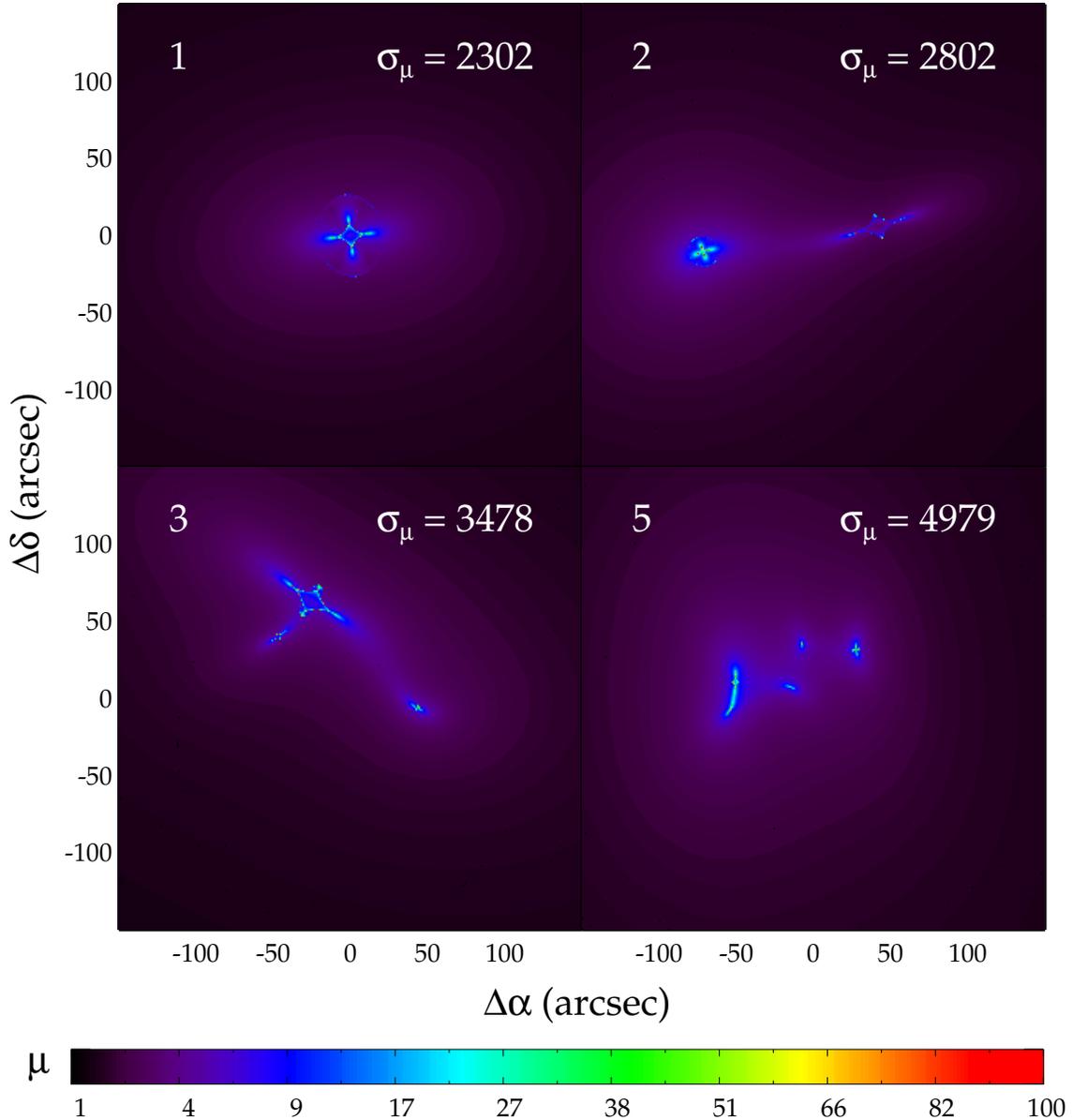}
\caption{Sample $z = 10$ source plane magnification maps over a $300\arcsec \times 300\arcsec$ field for lines of sight ranked near the top decile in $\sigma_{\mu}$ with a single halo (top left), two halos (top right), three halos (bottom left), and five halos (bottom right).  The colors correspond to the magnification of the single most magnified image of a point source at each position on the map.  The color map is logarithmic in the magnification.  The values of $\sigma_{\mu}$ (arcsec$^{2}$) for these particular configurations are indicated in the top right corner of each panel.  The increase in $\sigma_{\mu}$ as the number of halos increases is similar to the result shown in Figure~\ref{fig:magmap}.   \label{fig:magmap_top}}
\end{figure*}

\begin{figure*}
\centering
\plotone{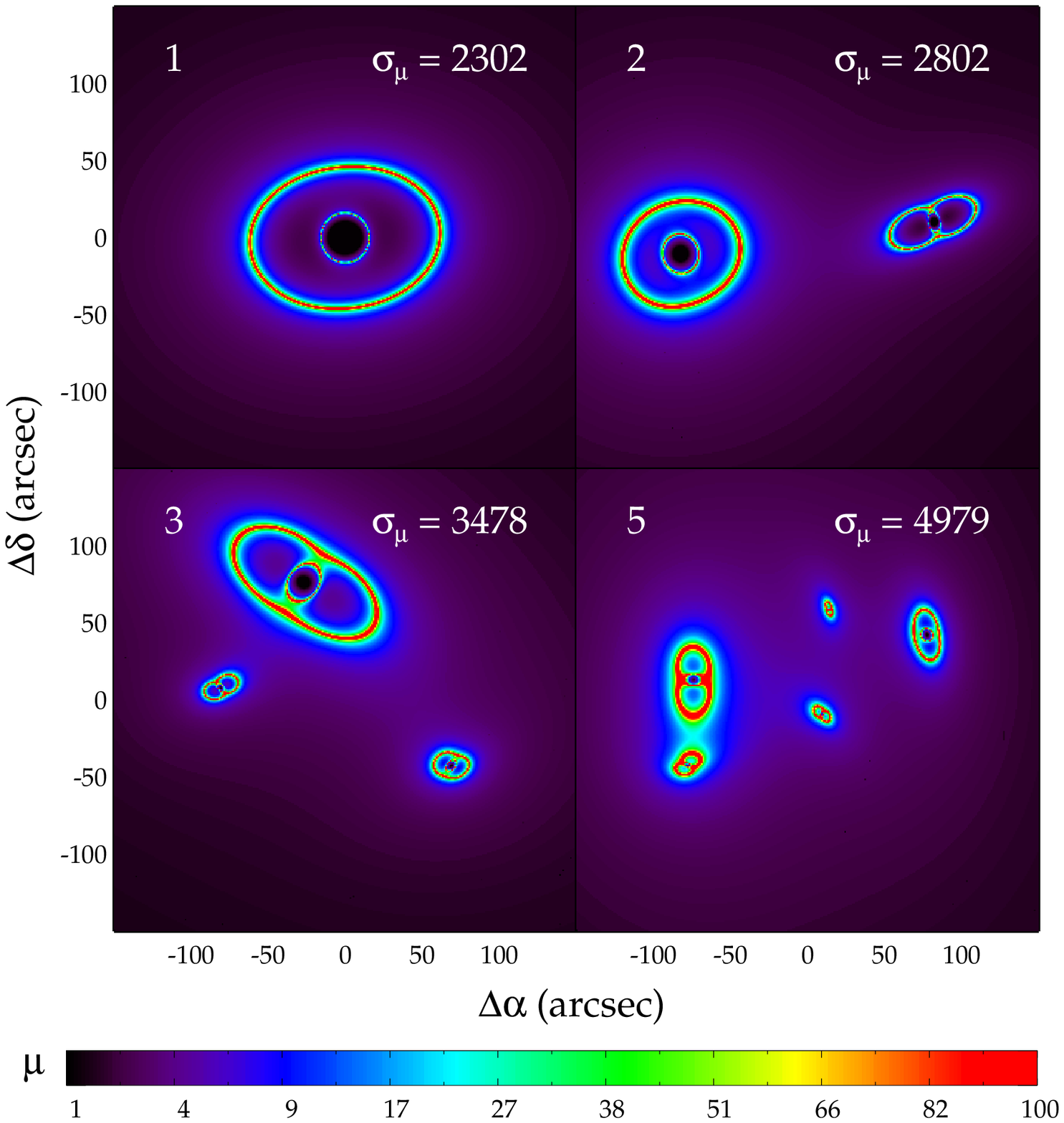}
\caption{Same as Figure~\ref{fig:magmap_top} but the maps in the image plane are now shown.  The magnification maps shown here correspond to the same mass distributions as in Figure~\ref{fig:magmap_top}.  \label{fig:magmapimg_top}}
\end{figure*}

We generate a sample of fields containing a single halo as a control sample with which to compare the multi-halo mass distributions.  These fields contain a single NFW halo with a virial mass within 1\% of our fiducial mass of $ 2 \times 10^{15} M_{\odot}$ with properties described in \S~\ref{subsec:halos}.  A source-plane magnification map for a typical one of our Monte Carlo trials is shown in the top left panel of Figure~\ref{fig:magmap}, with the corresponding image-plane map shown in the top left panel of Figure~\ref{fig:magmapimg}.  Cross-sections are evaluated in the source plane.  We will also investigate the properties of the best mass configurations that are ranked within the top 10\% by $\sigma_{\mu}$, so we plot analogous maps in the top left panels of Figures~\ref{fig:magmap_top} and~\ref{fig:magmapimg_top} for one such configuration.

\subsection{Two Halos} \label{subsec:2halo}
Our first multi-halo test is to look at lines of sight with a total mass of $2 \times 10^{15} M_{\odot}$ divided between two separate cluster-scale halos.  Sample magnification maps for one of our typical trials are shown in the top right panels of Figures~\ref{fig:magmap} and~\ref{fig:magmapimg}, with the same plots for a top 10\% trial show in the top right panels of Figures~\ref{fig:magmap_top} and~\ref{fig:magmapimg_top}.  Although the total mass in the beam remains fixed to within 1\%, we can quantify the differences in their lensing properties compared to the single-halo beams of similar total mass, as well as investigate the effects of varying other properties such as the halos' angular separation, relative mass ratio, and redshift distribution.

Although we account for the halo mass function and its redshift evolution when assigning relative halo properties in our analysis, we are, by construction, forcing lines of sight to have both a fixed number of halos and a fixed amount of total mass, not accounting for the likelihood of finding a high-mass beam containing multiple halos projected nearby on the sky.  \citet{oguri2009} find that chance superpositions of two cluster halos that together create a large Einstein radius are rare compared to single clusters with similar lensing properties.  However, we are by design investigating the {\it properties} of lines of sight that contain rare mass configurations that are favorable to lensing.  We address the likelihood of these chance projections existing in the universe in \S~\ref{subsec:likelihood}.  The frequency of such fields will be analyzed in more detail using cosmological simulations in a future work (French et al. in preparation).  We note, however, that we have identified several such lines of sight in the SDSS but will leave analyses of their properties to future work (Ammons et al. in preparation).

\subsection{More Than Two Halos} \label{subsec:3+halo}
We generalize the tests performed in \S~\ref{subsec:2halo} to allow for the total mass in the beam to be redistributed among three or more halos projected along the LOS.  For a trial with $N$ distinct halos, we draw $N$ halos until we find a beam with a total integrated mass within 1\% of our fiducial single halo.  We test lines of sight containing $N = 3$ and $N = 5$ halos.  Sample magnification maps for typical 3-halo and 5-halo cases are shown in the bottom row of Figures~\ref{fig:magmap} and~\ref{fig:magmapimg}, with a top 10\% case shown in the bottom row of Figures~\ref{fig:magmap_top} and~\ref{fig:magmapimg_top}.  Further trials of beams containing even more distinct halos are possible, but calculating their lensing properties becomes computationally intensive.  Since these lines of sight are increasingly rare in the real universe as more cluster-scale halos are added, we do not explore beams with more than five halos.

\section{RESULTS} \label{sec:results}
We present the results of our analysis of the lensing properties of the single-halo and multi-halo lines of sight and look into the physical properties driving the differences between them.  In addition, we assess the frequency of various properties of the mass configuration and investigate the parameters that lead to the very best lensing configurations, independent of frequency.

\subsection{The Effect of Multiple Halos On Lensing} \label{subsec:multihalo_results}
\begin{figure}
\centering
\plotone{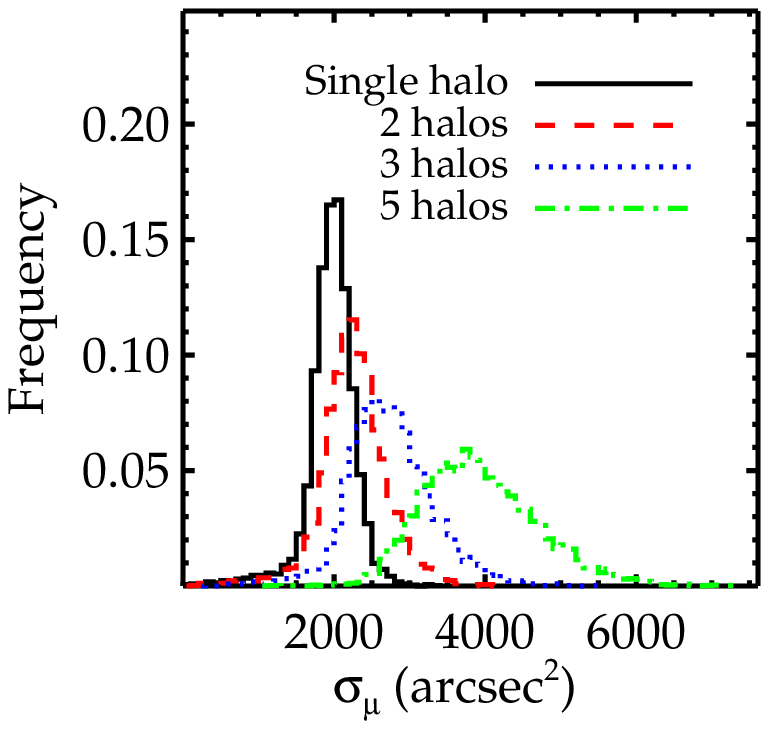}
\caption{Histogram showing comparisons of $\sigma_{\mu}$ for the single halo lines of sight to those of the multi-halo lines of sight from our Monte Carlo analysis.  5000 beams of each configuration were generated for each case.  The beams containing multiple halos have, on average, larger cross sections than the single-halo beams.  \label{fig:beamhist}}
\end{figure}

The results of our Monte Carlo analysis are presented in Figure~\ref{fig:beamhist}, where we show the distribution of $\sigma_{\mu}$ for all 5000 trials of each LOS mass distribution.  These results indicate that the multi-halo beams are, on average, better at magnifying high-redshift galaxies into detectability than single-halo beams of similar integrated mass.  Furthermore, this effect becomes more prominent as the mass is redistributed over more distinct halos.  The statistics of the distributions are quantified in Table~\ref{tab:hist}.

The typical improvement in the average lensing cross section of the 2-halo beams over the single-halo beams of equivalent mass is roughly $\sim 15$\%.  The improvement for the 3-halo and 5-halo beams are roughly $\sim 40$\% and $\sim 100$\%, respectively.  These gains are specific to our chosen fiducial mass and parameter ranges and will differ for various mass and separation range, but the overall effect remains similar.

\begin{table*}
\caption{Lensing Properties \label{tab:hist}}
\begin{ruledtabular}
\begin{tabular}{cccc}
Number of Halos &
$\langle \sigma_{\mu} \rangle$\tablenotemark{a} (arcsec$^{2}$) &
Median($\sigma_{\mu}$) (arcsec$^{2}$) &
Top 10\% Decile in $\sigma_{\mu}$ (arcsec$^{2}$)
\\
\tableline
1 &
1965 $\pm$ 5 &
1990 &
2300
\\
2 &
2268 $\pm$ 6 &
2262 &
2799
\\
3 &
2747 $\pm$ 8 &
2708 &
3470
\\
5 &
3939 $\pm$ 11 &
3859 &
4972
\\
2 (individual)\tablenotemark{b} &
1835 $\pm$ 5 &
1868 &
2190
\\
\end{tabular}
\end{ruledtabular}
\tablenotetext{1}{Errors given are standard errors on the mean.}
\tablenotetext{2}{Each of the beams in the 2-halo trials were split into individual halos and the lensing cross sections of each were summed.}
\end{table*}

\begin{figure}
\centering
\plotone{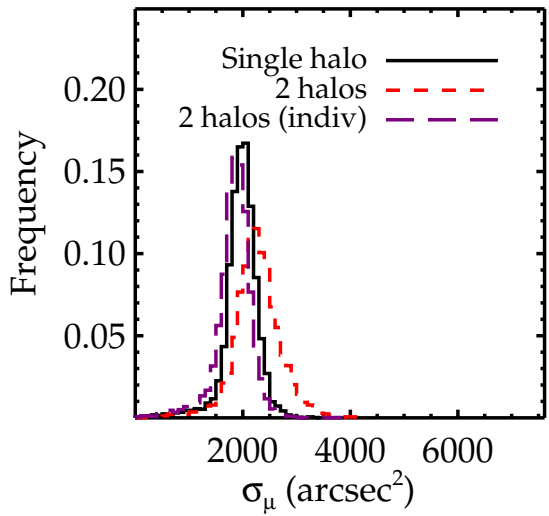}
\caption{Same as Figure~\ref{fig:beamhist}, but comparing only the single-halo and 2-halo lines of sight, as well as the sum of the quantities in each trial determined from each of the individual halos separately in the 2-halo cases.  The beams containing two halos have, on average, larger cross sections compared to the two halos considered separately. \label{fig:indivcomp}}
\end{figure}

It may be the case that splitting a single massive halo into multiple smaller halos increases the lensing cross section simply by virtue of covering a larger area in the sky.  It is not immediately obvious that it is the combined effect of multiple halos nearby in projection that is creating this effect.  To check this, we run a test where for each 2-halo trial, the two halos are analyzed as if they were each individual clusters, and their total lensing cross sections are summed.  Figure~\ref{fig:indivcomp} shows histograms of $\sigma_{\mu}$ for these trials compared to both the single-halo beams and the 2-halo beams.  Table~\ref{tab:hist} also shows the mean and median values of the cross section for these trials.  The results show that simply dividing a single halo into two smaller independent halos does not result in the same gain that is seen when it is split into two halos acting in combination along the same line of sight.  In fact, doing so actually decreases the lensing power of the mass distribution.  It is the combination of the two halos in projection acting together that results in the increase seen in Figure~\ref{fig:beamhist}.  We discuss this effect further in \S~\ref{subsec:proximity}.

Additionally, one can compare the gain in cross section due to configuration to the gain due to an increase in total mass, as shown in Figure~\ref{fig:sigmamass}.  While this basic comparison only represents the scaling for one simplified scenario, a general sense of how much the configuration can affect the cross section compared to a change in the total integrated mass can be inferred.  For example, comparing the mean cross section in the single halo case to that of the 5-halo case shows an increase of a factor of $\sim$2, which is roughly comparable to changing the total mass of the single halo by a factor of $\sim$2.  This demonstrates that the effect of configuration is non-negligible compared to the effect of total mass, and may even be comparable over this range in total mass.

\subsection{Halo Properties Important for Lensing} \label{subsec:params}
It is important to determine what parameters drive correlations between the physical properties of the multi-halo mass distributions and their lensing properties.  We focus on the 2-halo beams for this discussion, as many of the relevant physical properties of these lines of sight can be represented by a single quantity, and 2-halo beams will be more common than lines of sight containing three or more cluster-scale halos.

\begin{figure*}
\centering
\plotone{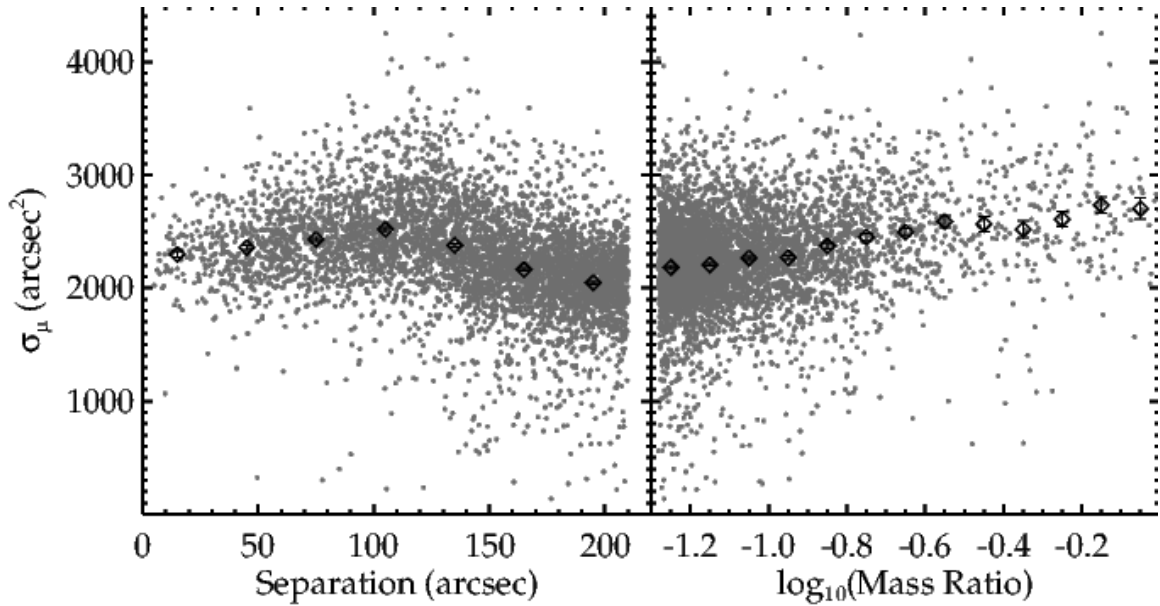}
\caption{$\sigma_{\mu}$ as a function of angular separation (left) and mass ratio (right) of the halos in the 2-halo trials.  The grey points represent individual trials, and the black points are averages in each bin.  The error bars represent standard errors on the mean.  {\it Left:} $\sigma_{\mu}$ rises with separation and peaks around $\sim$100\arcsec, then drops off at larger separations.  This is consistent with the understanding that at large separations, the halos approach the limit where the regions around them can be treated as independent lensing fields.  {\it Right:} The mass ratio is defined as the mass of the smaller halo divided by the mass of the larger halo.  Mass ratios closer to unity produce larger lensing cross sections, with $\sigma_{\mu}$ generally decreasing with smaller mass ratio.  At low mass ratio, we approach the limit of the single-halo fields, which have lower cross sections on average than the 2-halo case. \label{fig:sigmamusepmr}}
\end{figure*}

The results of \S~\ref{subsec:multihalo_results} show that the lines of sight comprising multiple halos tend to be better lenses than the total of each of their constituent halos considered individually, assuming fixed total mass.  This suggests that for a given set of halos, there is an optimal configuration in the plane of the sky that maximizes the lensing signal.  For the 2-halo case, this corresponds to an optimal angular offset given two particular halos.  In the left panel of Figure~\ref{fig:sigmamusepmr}, we plot $\sigma_{\mu}$ for each trial as a function of the angular separation between the two halos.  Due to the angular positions of the halos relative to one another, there is a higher overall frequency of beams at larger projected separations, as we would expect if the halo positions on the sky are uncorrelated.  Despite the fact that this visualization marginalizes over all other parameters, it is clear that $\sigma_{\mu}$ rises gradually, peaks at a separation of $\sim$100\arcsec, and then decreases at higher separations.

The dropoff at large separation appears to be sensible in the context of the results of Figure~\ref{fig:indivcomp}, since  we are approaching the limit where the regions around each halo can be treated as independent fields.  The peak at nonzero separation is more interesting, and can be explained by the distribution of surface mass density when multiple mass components are spread apart.  We discuss the interpretation and implications of this result further in \S~\ref{subsec:proximity}.

We perform a similar test to examine the effect of varying the relative masses of the halos in the beam.  The right panel of Figure~\ref{fig:sigmamusepmr} shows $\sigma_{\mu}$ as a function of the log of the ratio of the two halo masses (defined such that the mass ratio is always $\leq 1$).  The trend shows that mass ratios closer to unity results in larger cross sections.  As the mass ratio becomes more extreme, we are approaching the limit of the single-halo case, so this result makes sense intuitively.  As an additional check of these results, we run 2-halo trials where we both increase and decrease the minimum mass allowed for our halos (see \S~\ref{subsec:uncertainties}).  Both of these tests show the same trend of $\sigma_{\mu}$ with mass ratio, and the mean $\sigma_{\mu}$ across all trials increases as the minimum mass (and thus, the mean mass ratio) increases, supporting our conclusion.

We note that Figure~\ref{fig:sigmamusepmr} indicates a large number of trials with extreme mass ratios where there is a single dominant halo.  Due to the steepness of the halo mass function above $10^{14} M_{\odot}$ within our redshift range, drawing $N$ halo masses that add up to our fiducial mass is much more likely to select one massive halo along with $N-1$ halos close to our minimum mass of $10^{14} M_{\odot}$ than it is to select $N$ similarly massive halos, which results in this skewed distribution.  One may wonder in this case whether the lensing properties of the multi-halo beams are dominated by the physical properties of the more massive halo.  To check this, we look at how the properties of the single-halo beams affect the lensing cross section and compare any trends or features to those in the multi-halo beams where only the properties of the more massive halo are considered.

\begin{figure*}
\centering
\plotone{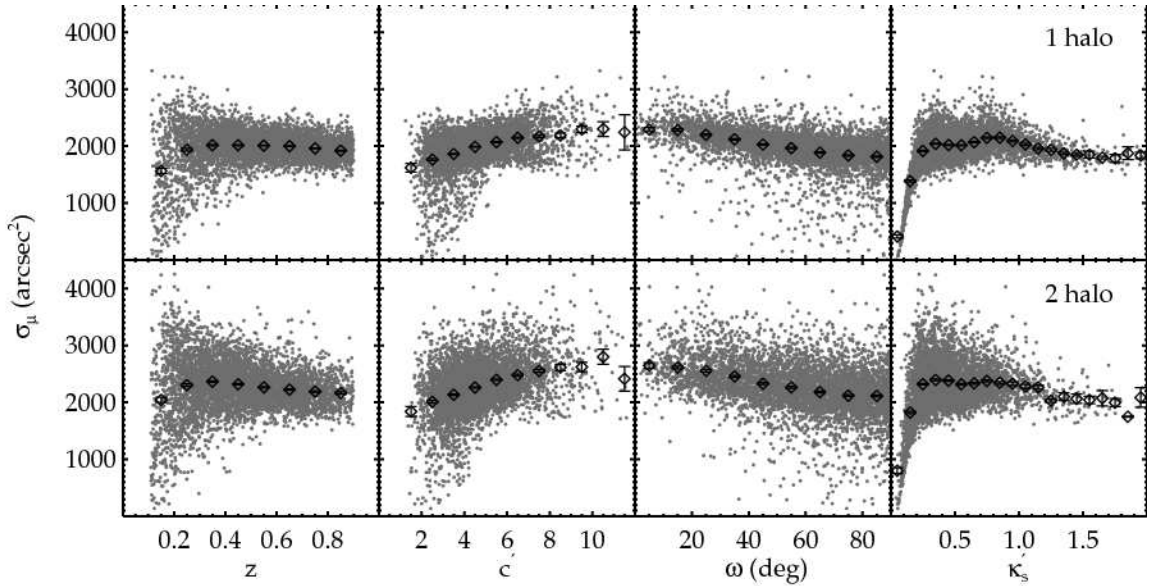}
\caption{$\sigma_{\mu}$ as a function of (from left to right) redshift, projected concentration, angle between the major axis and the line of sight (denoted $\omega$), and projected scaled NFW surface density $\kappa_{s}^{\prime}$ of the halos in the single-halo sample (top) and the more massive halo in each trial in the 2-halo sample (bottom).  Other parameters, including projected ellipticity and triaxiality, are not plotted because they do not show a statistically significant trend.  $\sigma_{\mu}$ is larger in general at redshifts $z \gtrsim 0.3$ compared to lower redshifts, but is fairly constant with redshift on average beyond that point.  $\sigma_{\mu}$ also increases with larger projected halo concentration and decreasing $\omega$.  $\sigma_{\mu}$ rises sharply with $\kappa_{s}^{\prime}$ at low values, then plateaus due to the way we have defined the cross section in our analysis (see text for details).  The results in the two-halo case show that the properties of the more massive halo affect the cross section in the same way as the single-halo case because most trials have a single dominant halo.  We note that $c^{\prime}$ and $\kappa_{s}^{\prime}$ are the projected concentration and scaled surface density after accounting for triaxiality and projection effects (see Appendix~\ref{app:projection}).\label{fig:onehaloparamsmulti}}
\end{figure*}

In the top row of Figure~\ref{fig:onehaloparamsmulti}, we plot $\sigma_{\mu}$ for the single-halo lines of sight as a function of various fundamental parameters.  We see that halos at redshifts higher than $z \sim 0.3$ tend to be better lenses, which is due to the fact that the critical density for lensing, $\Sigma_{c}$, has a broad minimum centered at $z \sim 0.8$, resulting in an optimal geometry for lensing $z = 10$ sources.  We also see that halos with higher projected concentration, $c^{\prime}$, tend to have larger cross sections, which is expected due to the increase in surface density that this implies.  $c^{\prime}$ is determined both by the intrinsic concentration of the halo, as well as density enhancements due to triaxiality and projection effects (see Appendix~\ref{app:projection}).  This results in additional scatter in $c^{\prime}$, with halos that have their major axis aligned with the line of sight tending to have larger concentrations, and therefore larger cross sections.  We use $\omega$ to denote this angle, which is defined to be zero when there is perfect alignment between the major axis and line of sight direction.  The effect is small in comparison to the intrinsic concentration of the halo, but we see in the third panel of Figure~\ref{fig:onehaloparamsmulti} that it is not negligible.

Other physical parameters do not show a statistically significant correlation with $\sigma_{\mu}$, so we do not plot these quantities.  These parameters include projected ellipticity and triaxiality, as well as relative ellipticity position angles between both halos or between the more massive halo and the angle to its companion.  Given that there is a trend with $\omega$, one might have expected a possible trend with halo triaxiality.  However, because $\omega$ is defined by the direction of the major axis, regardless of halo shape, this is not the case.  Likewise, multiple combinations of axis ratios and viewing angles can result in similar projected ellipticities, so there are no obvious dependencies on ellipticity alone.

We also plot $\sigma_{\mu}$ as a function of the projected scaled surface density for NFW halos, $\kappa_{s}^{\prime}$ (see Appendix~\ref{app:projection}), in the upper rightmost panel of Figure~\ref{fig:onehaloparamsmulti}.  $\kappa_{s}^{\prime}$ is dependent on other fundamental parameters, but we show it here because this relation shows an interesting shape.  Very low values of $\kappa_{s}^{\prime}$ correspond to small $\sigma_{\mu}$, which occurs when the halo is at a low redshift such that it has a large angular size and therefore a small surface density.  One might expect that the cross section would monotonically increase as the surface density increases, but instead, $\sigma_{\mu}$ reaches a point where it plateaus and eventually turns over.  This unexpected behavior has to do with our definition of the cross section.  As $\kappa_{s}^{\prime}$ increases, the area in the source plane interior to the inner caustic grows.  Any sources that fall within the caustic get multiply imaged, but the maximum magnification of any individual image is sometimes lower than our threshold magnification, $\mu_{t}$.  Therefore, this area does not count toward the cross section as we have defined it.  Obviously, different magnification thresholds will change the turnover point, but adding more surface density in general will increase this area but will {\it not} increase $\sigma_{\mu}$ once a certain point has been reached.  We note that the traditional strong lensing cross section, defined as the area in the source plane where a source is multiply-imaged, does increase monotonically with $\kappa_{s}^{\prime}$ as expected.

We examine whether these results also hold true for the multi-halo distributions where only the properties of the more massive halo are considered.  In the bottom row of Figure~\ref{fig:onehaloparamsmulti}, we plot $\sigma_{\mu}$ for the 2-halo lines of sight as a function of the same parameters for the more massive halo in each trial.  The properties of the dominant halo affect the cross section in a similar way as in the single-halo case, as we would expect when one halo dominates the mass distribution.  While 1:1 mass ratios generally produce higher $\sigma_{\mu}$, even the far more numerous $\geq$ 10:1 mass ratios can generate high $\sigma_{\mu}$.  One implication of Figure~\ref{fig:sigmamusepmr} is that in studies of known strong lensing clusters where LOS structures are typically ignored, any massive structures in the beam that go undetected are more likely to be small clusters rather than one with a mass equivalent to the main lens.  As such, not accounting for these structures is unlikely to result in a large error on the cross section of these beams.  However, the effects of these structures on the derived 2-D magnification maps, which we do not analyze here, may result in systematic uncertainties that will cause the magnification errors in the field to be underestimated.

\subsection{Properties of the Best Lensing Configurations} \label{subsec:top10}
In addition to identifying the general trends of lensing cross section with various parameters, we also examine how the mass configurations resulting in the very best $\sigma_{\mu}$ values behave.  The results of Figures~\ref{fig:sigmamusepmr} and~\ref{fig:onehaloparamsmulti} show considerable scatter about the mean relation, so there may be parameters that can cause particular mass configurations to scatter to a high cross section.  These mass configurations, though likely very rare in the universe, could potentially be the very best lines of sight for lensing high-redshift galaxies.

\begin{figure*}
\centering
\plotone{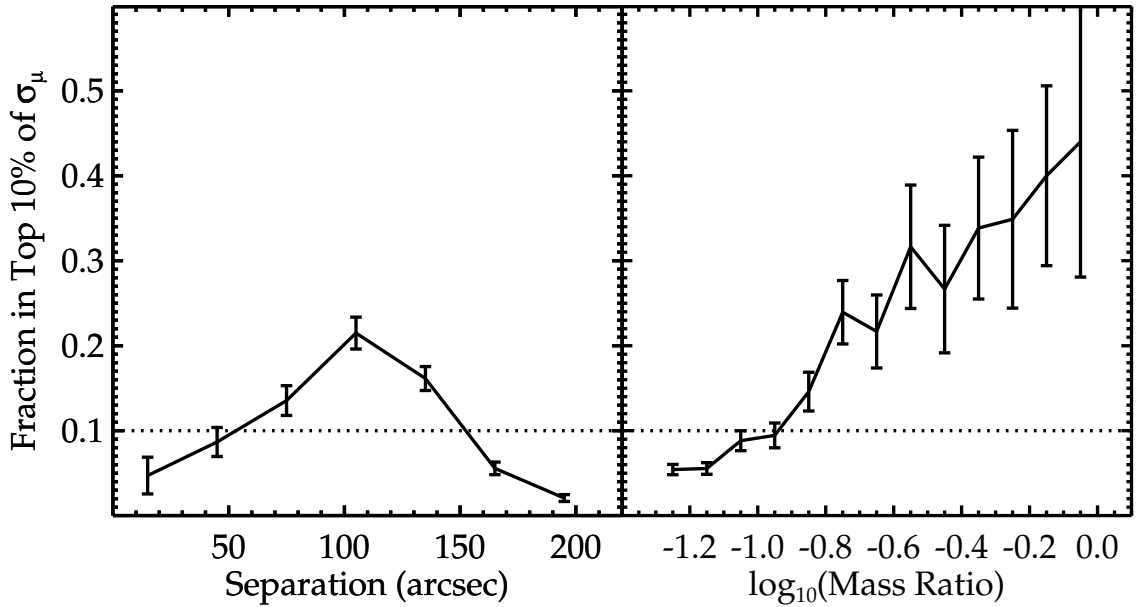}
\caption{Fraction of 2-halo beams in each separation (left) and mass ratio (right) bin ranked in the top 10\% of the entire sample by $\sigma_{\mu}$.  The dotted line represents a frequency of 10\%, which would be expected if there were no excess of top $\sigma_{\mu}$ configurations in a particular bin.  These relations match what would be expected based on the mean relations $-$ there is an optimal nonzero angular separation where the best mass configurations lie, and mass ratios closer to unity generally produce good lensing configurations.  The error bars are larger in bins with fewer samples.  \label{fig:top10}}
\end{figure*}

We examine the frequency of 2-halo lines of sight that are ranked in the top 10\% when sorted by $\sigma_{\mu}$ as a function of various parameters.  The values of $\sigma_{\mu}$ corresponding to the top 10\% decile are given in Table~\ref{tab:hist}.  In Figure~\ref{fig:top10}, we plot the fraction of trials that fall within the top 10\% in $\sigma_{\mu}$ in each bin for both projected separation and mass ratio.  This plot represents the excess (or deficit) of high $\sigma_{\mu}$ mass configurations in each parameter bin.  By plotting the results in this manner, we can see if the parameter ranges that generate the very best mass configurations for lensing are similar to what would be expected from the mean trends for the full sample.  In addition, this visualization removes the effects of our priors on the parameter being plotted such that we are not inherently biased toward a region of that parameter space that was just sampled more frequently and would therefore tend to have a larger raw number of configurations that fell in the top 10\% due to scatter.  We are effectively examining what values of a particular parameter result in the best lensing configurations regardless of their actual likelihood, provided that likelihood is not vanishingly small.

We see that the angular separation and mass ratio dependencies are consistent with the mean results, with the separation bin around $\sim$100\arcsec~and mass ratio bins near unity containing a larger fraction of top 10\% configurations.  The dependence on angular separation demonstrates that our choice of a $3\farcm5$ maximum separation is justified, as the very best mass configurations tend to have separations smaller than this.

\begin{figure*}
\centering
\plotone{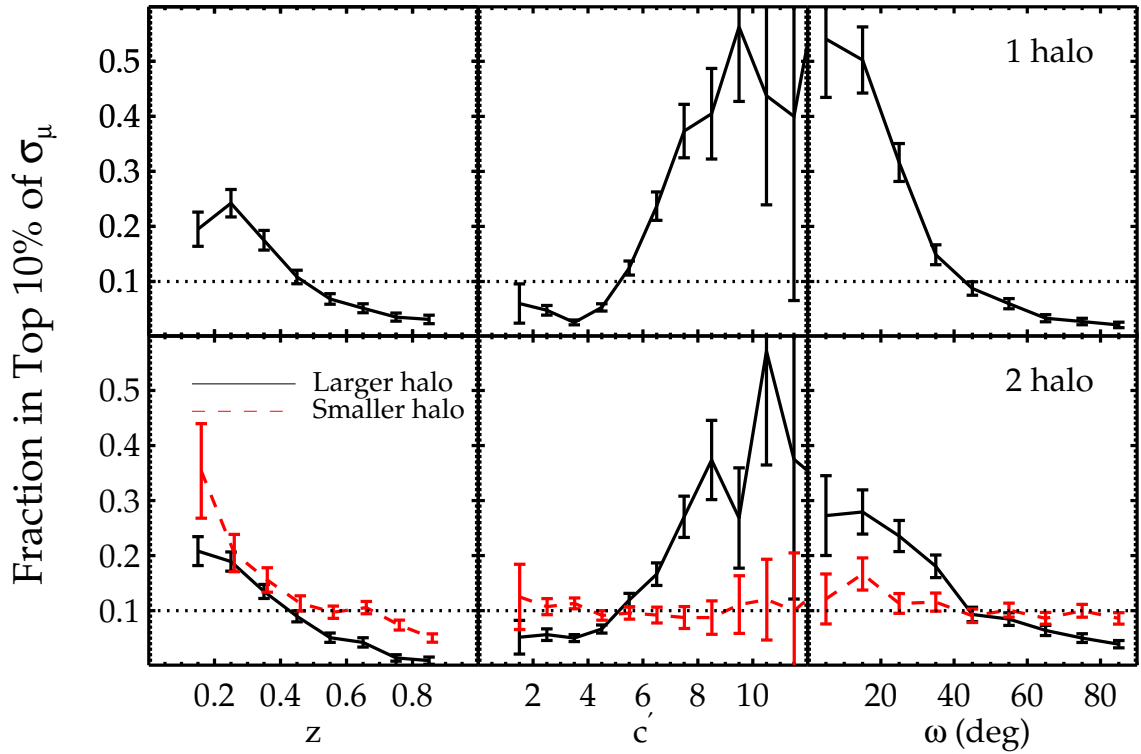}
\caption{{\it Top:} Fraction of single-halo beams in each redshift (left), concentration (center), and orientation angle (right) bin ranked in the top 10\% in $\sigma_{\mu}$.  {\it Bottom:} Fraction of 2-halo beams in each bin of the same parameters for the more massive (black solid line) and less massive (red dashed line) halo in each trial that is ranked in the top 10\% in $\sigma_{\mu}$.  The dotted line represents a frequency of 10\%, which would be expected if there were no excess of top $\sigma_{\mu}$ configurations in a particular bin.  This visualization eliminates the effect of the frequency of certain mass configurations in the parameter being plotted on the x-axis.  The concentration and $\omega$ relations for the single halo and more massive halo in the 2-halo case match what would be expected from the mean relations in Figure~\ref{fig:onehaloparamsmulti}.  Low redshift bins ($z \sim 0.2$) contain a larger fraction of top 10\% mass configurations, despite the mean trend tending to disfavor low redshift.  The physical properties of the less massive halo in the 2-halo case are generally unimportant since there is usually a single dominant halo, even in the top 10\% trials.  There is a preference for the lower mass halo to be at low redshift, but this arises at least in part from a selection effect where such configurations are also likely to have the higher mass halo at low redshift, and the solid line indicates that having the higher mass halo at low redshift is more likely to result in a top 10\% $\sigma_{\mu}$ configuration.  \label{fig:top10_1halo}}
\end{figure*}

We make similar plots for the physical properties of the halos in the single-halo case in the top row of Figure~\ref{fig:top10_1halo}.  Interestingly, the frequency of top 10\% mass configurations peaks at low redshifts ($z \sim 0.2-0.3$) and decreases with increasing redshift.  This is in contrast with the mean trend for the entire sample, which shows lower $\sigma_{\mu}$ at low redshift with a subsequent rise and plateau at higher redshift.  Revisiting the upper left panel of Figure~\ref{fig:onehaloparamsmulti}, there is a subset of trials with the very highest cross sections located at low redshifts, but there is also a large subset of trials in these bins that have very low $\sigma_{\mu}$ values.  What is happening is that at these low redshifts, the NFW scale radius, $r_{s}$, becomes large in angular size, even though its physical scale radius may be comparable to higher redshift halos of similar mass.  This implies a decrease in the surface density, which decreases the lensing strength of the halo.  In addition, the critical density, $\Sigma_{c}$, is larger at lower redshift, further decreasing the lensing strength.  However, halos that have an unusually high concentration at these redshifts will have a large surface density over a large angular area, creating a favorable lens.  Indeed, an inspection of the high-cross section trials at low redshift reveals a disproportionate amount of halos with high concentration.  In general, a halo at low-redshift will have an unfavorable lensing geometry compared to a similar halo at higher redshift, but in certain rare cases, the concentration can be high enough that the halo becomes a very good lens.  An example is Abell 1689, which has a redshift of $z = 0.18$, but has an unusually high concentration compared to expectations from cosmological simulations \citep{broadhurst2005,halkola2006,zekser2006,limousin2007,umetsu2008,coe2010,sereno2011}, making it one of the best lensing clusters known.

The other physical parameters that we found relevant, namely the projected concentration and angle between the major axis and the line of sight, behave in the way expected from the mean trends $-$ higher concentration and better alignment with the LOS results in better lenses, and thus a higher chance of attaining a very high cross section.

We make similar plots in the bottom row of Figure~\ref{fig:top10_1halo}, but we now plot the relative fractions for the 2-halo beams as a function of the physical parameters of both the more massive and less massive halo in each trial.  If the properties of the more massive halo dominate the effect on the lensing properties, as was suggested by Figure~\ref{fig:onehaloparamsmulti}, then we would expect to see similar trends with the properties of the more massive halo, but essentially no trend with those of the less massive halo.  This is true for the concentration and for $\omega$, as the trends with the more massive halo are similar to those in the single-halo case, and the trends with the less massive halo are flat.  However, there is a trend with redshift of the less massive halo that is qualitatively similar to that with both the redshift of the high mass halo and of the single-halo case.  This result arises at least in part from the shape and evolution of the halo mass function.  In roughly 3/4 of our 2-halo trials, the more massive halo in a pair is at lower redshift than the less massive one.  In other words, when the less massive halo is at a low redshift, it is more likely for the other, more massive halo, to also be at low redshift.  Furthermore, such configurations tend to have less extreme mass ratios.  Since low redshift halos show a higher frequency of top 10\% $\sigma_{\mu}$ values, it appears that the redshift of the less massive halo is having an effect, but in fact, this trend is primarily an artifact of the mass function and our requirement that the lines of sight have a fixed total mass rather than a real lensing effect.

The dominance of a single massive halo in most of the 2-halo case initially appears to be inconsistent with the trend showing that more equal mass ratios are more likely to result in a good configuration.  In fact, the fractions shown in the right panel of Figure~\ref{fig:top10} indicate that there are still many excellent mass configurations that do not follow the trend, especially considering that the bins representing the most extreme mass ratios contain far more trials compared to the bins of more equal mass ratio.  The point is that none of these parameters are {\it required} to be optimized in order to produce a good mass configuration.  Given a fixed total mass, there are certainly parameter values that increase the likelihood of this occurring, but there are multiple ways and combinations of parameters that can result in a large cross section.

\section{DISCUSSION} \label{sec:discussion}

\subsection{Lensing Interactions Between Halos} \label{subsec:proximity}
Our results show that there is a gain in the lensing cross-section when a massive halo is split into two or more cluster-scale halos projected close on the sky.  However, the results presented in \S~\ref{subsec:multihalo_results} indicate that it is not simply the lensing properties of the individual halos that sum up to produce a more powerful lens than a single halo of equivalent total mass.  In fact, the values shown in Table~\ref{tab:hist} and Figure~\ref{fig:indivcomp} demonstrate the opposite effect.  Instead, the gain in cross-section for the multi-halo beams over the single-halo case of equivalent mass arises from interactions between the lens potentials of the halos, even if the halos are not physically interacting.  The results presented in this paper are specific to our chosen fiducial mass and range of halo separations, but some general statements can be inferred.

There is an optimal separation ($\sim 100$\arcsec~for our chosen fiducial total mass) between halos that maximizes the cross-section (Figure~\ref{fig:top10}).  One can understand this conceptually as follows.  Much of the contribution to $\sigma_{\mu}$ comes from regions of the source plane that are outside the strong lensing caustics but still significantly magnified (see Figures~\ref{fig:magmap} and~\ref{fig:magmap_top}).  Outside the caustics, the magnification produced by a given halo generally decreases with distance from the halo.  When two or more halos are placed in proximity to each other on the sky (even if they are at different redshifts), their magnification regions in the source plane can overlap and create areas where no halo individually exceeds our chosen magnification threshold but the joint lensing effects boost $\mu$ above the threshold.  While this notion of overlap is useful heuristically, recall from \S~\ref{subsec:lensing} that the net magnification is {\it not} simply the sum of the magnifications from individual halos.  As indicated by Equations~\ref{eq:mag}--\ref{eq:sheartens}, we need to consider not only the overlap of the halos' surface density profiles (the $\kappa$ terms), but also the combined effects from shear (the $\gamma$ terms).  In other words, the ``lensing interaction'' between the halos is mediated by both convergence and shear.

As the halos' angular separation decreases, the boost from the lensing interaction begins to increase.  At some point, however, the separation becomes small enough that the caustics begin to merge on the sky, which may increase the area where the magnification is very large (e.g., $\mu \gtrsim 10$--100) at the cost of reducing regions where the magnification is more modest but still significant (e.g., $\mu \sim 3$--10).  In other words, finite separations can maximize the area that produces modest magnification, whereas smaller separations could be more relevant if we were interested in larger magnifications (see Appendix~\ref{app:muthreshold}).

A related effect explains why having multiple halos helps.  As a beam containing a fixed total mass is split into multiple halos, each halo's average mass decreases and so its magnification region shrinks.  At the same time, the average projected separation between halos also decreases (when the beam size is fixed).  For the beam sizes and masses we have considered, the decrease in $\sigma_\mu$ from the dilution of mass is more than offset by the increase in $\sigma_\mu$ from the proximity of halos, so the cross sections tend to increase when we consider beams with more halos but the same total mass.  The \'{e}tendue of the mass configuration, defined as magnification over a large area, is increased as the mass is dispersed, provided that it is not spread out so thin that regions in between the halos drop below the magnification of interest.

We have examined multi-halo trials where we have changed the total integrated mass, and while the optimal offset may change, our general conclusion that splitting the mass into multiple halos holds true, provided that the typical angular separation between halos is roughly comparable to the size of their region of influence.

\subsection{Additional Sources of Uncertainty} \label{subsec:uncertainties}
We have generated mass configurations that are physically motivated by the relative frequencies and properties of cluster-scale halos as determined from simulations, with the exception that we have constrained these lines of sight to have a large total mass to represent powerful lensing configurations.  Although our treatment of these massive halos accounts for much of the diversity seen among such halos with regard to properties such as redshift and mass distribution, concentration, and shape/orientation, among other things, there are some details that we have not considered.

While we allow for multiple dark matter halos at similar redshift in our models, we do not account for substructure in our halos.  This includes the effects of galaxies and higher-order details in the halo models \citep[e.g. misaligned isodensity contours, mass-shape correlations;][]{schneider2011}.  \citet{hilbert2008} find that the lensing cross sections for halos more massive than $10^{14} M_{\odot}$ increases by at most 40\% for a source redshift of $z = 5.7$ and is more significant for lower masses when the effects of galaxies are included, although their definition of cross section is somewhat different from ours.  The galaxies will affect both the single-halo and multi-halo cases, and in fact could have a larger impact on the multi-halo fields where the typical halo mass is less than the fiducial mass, so our assumption of no substructure due to a galaxy component may be conservative.  The effects of baryonic cooling can increase the lensing cross section of these halos \citep{rozo2008}, but this should affect both single-halo and multi-halo lines of sight in a similar fashion.  We also assume that our halos follow an NFW density profile, and deviations from NFW could alter the calculated cross sections.  However, as noted earlier, this assumption makes our analysis compatible with previous work and is suggested by some observational results \citep{kneib2003,katgert2004,comerford2006,shu2008}.

Although we allow for the possibility of halos being at the same redshift, we do not account for physical interactions between halos that will alter the dark matter distribution, such as mutual tidal effects, density enhancements in the baryonic components (e.g., the Bullet Cluster \citep{clowe2006} or Abell 2744 \citep{merten2011}), and multiple subclusters sharing a common dark matter halo.   These effects may lead to an overall underestimate of the lensing cross section, but should not preferentially affect a given multi-halo configuration.

We also do not account for smaller uncorrelated structures along the line of sight.  These low-mass structures, which include field galaxies and smaller groups, should have small effects on the lensing cross section of the main cluster(s) in these fields if the results of Figure~\ref{fig:sigmamusepmr} are extrapolated to lower mass ratios.  We do not expect such structures to be more prevalent in beams with fewer halos, so it will not affect our general conclusions about multi-halo fields.  We have required that the halos in our analysis have a minimum mass of $10^{14} M_{\odot}$, but less massive halos down to the group scale ($\sim 10^{13} M_{\odot}$) are more common and are more likely to be found in multi-halo configurations.  We run a test where we set the minimum halo mass in the 2-halo case to $5\times10^{13} M_{\odot}$ and find that $\sigma_{\mu}$ is still larger for these cases on average than for the single-halo case.  This also can be seen from the trend in mass ratio, where a larger difference between the two halo masses results in a smaller $\sigma_{\mu}$.

\subsection{Implications for Detection of High-Redshift Sources} \label{subsec:ndetect}
Gravitational lensing is effectively a mapping from a given area in the source plane to an area in the image plane, which aids in the detection of faint sources by increasing the solid angle subtended with no loss in surface brightness.  However, this magnification comes at the cost of a decreased area being probed in the source plane.  As a result, the gains achieved from lensing in terms of detecting high-redshift galaxies depends very strongly on the shape of the source galaxy luminosity function (LF).  The LF is typically assumed to be of the form first proposed by \citet{schechter1976}, which is parameterized by a characteristic magnitude $M^{*}$, a faint-end slope $\alpha$, and a normalization factor $\phi^{*}$ representing the number density at the characteristic luminosity.  For galaxies fainter than $M^{*}$, the LF is approximately a power law with slope $\alpha$.  For galaxies brighter than $M^{*}$, the LF declines exponentially with increasing luminosity.

Since the loss in area probed is inversely proportional to the lensing magnification, lensing increases the likelihood of detecting a galaxy at a given source redshift if the LF at that galaxy's luminosity is declining more steeply than a power law of index $-2$ (not accounting for finite source size effects; see Ammons et al. in preparation).  Therefore, lensing is mainly beneficial for detecting the most luminous sources \citep{broadhurst1995,santos2004,bouwens2009}, unless the faint-end slope is steep.  The shape of the LF is not well constrained at $z \gtrsim 7$, so it is not immediately obvious how $\sigma_{\mu}$ would translate to an enhancement in the detectability of high-redshift sources.

We have presented our results in terms of lensing cross-section or \'{e}tendue.  Observational studies of the high-redshift universe have a variety of science objectives that may require different observing strategies in terms of depth versus area or volume.  For example, given a fixed amount of observing time, constraining the LF faint-end slope requires a very deep observation, while constraints on the bright-end are optimized with shallower observations over a wider area.  Lensing will detect fainter sources than a similar blank field observation within the same integration time.  Here, we examine the specific case of detecting $z \sim 10$ galaxies and the extent to which multi-halo beams provide an increase in the frequency and depth of detected sources relative to blank fields and single lensing clusters.  

The detailed calculation depends on several variables, including the source size and ellipticity distribution, the instrument PSF, the depth and field of view of the observing program, and the source luminosity function.  The only observational constraint on the LF at $z \sim 10$ is based on $\sim 0.8$ detections \citep{bouwens2011a} with ultra-deep HUDF observations using {\it HST}/WFC3.  Because of these weak constraints, a calculation of the number of $z \sim 10$ detections that would be expected in fields containing multiple cluster-scale halos is extremely uncertain.  However, we can make some simplifying assumptions and estimate the number of detections at different luminosities relative to a blank field of equivalent size.

The enhanced detectability of high-redshift sources due to lensing does not scale linearly with magnification due to the effects of finite source size (Ammons et al. in preparation).  $z \gtrsim 7$ sources are marginally resolved in high-resolution space-based imaging \citep{barkana2000,ferguson2004,bouwens2008,oesch2010a} and become smaller at higher redshifts if a scaling of $R_{eff} \propto (1+z)^{-1.1}$ is assumed \citep{bouwens2006}.  For unresolved sources, the effects of lensing magnification can increase their solid angles to sizes larger than the detector PSF.  For a lensed source resolved in one, two, or neither directions on the sky, the fractional increase in its S/N compared to an unlensed source, or ``signal-to-noise boost'' (SNB), can scale differently with magnification in different regimes.  This effect must be considered when calculating the detectability of high-redshift sources.

We focus on the case of a hypothetical space-based search for high-redshift galaxies and assume a PSF full-width at half-maximum of 0.15\arcsec, approximately the size of the $HST$/WFC3 PSF in the $H_{160W}$ filter.  We assume a population of sources with a Gaussian shape at $z = 10$ with a normal distribution of effective radii with a mean of 0.12\arcsec~and standard deviation of 0.04\arcsec.  This angular size distribution is determined by applying a size evolution of $(1+z)^{-1.1}$ \citep{bouwens2006} to the half-light radius distribution of $z \sim 7$ sources, which is $0.75 \pm 0.23$ kpc for galaxies with luminosities in the range $0.3L^{*} < L <L^{*}$ \citep{oesch2010a}.  We assume a normal distribution of intrinsic major-to-minor axis ratios with a mean of 1.2 and a standard deviation of 0.3, which can be folded into the SNB calculation using the formalism in \citet{keeton2001b}.  The effect of intrinsic shape on the detectability of a source is much smaller than the effect of source size.

The field of view and observational depth will also affect the number of detected sources.  We calculate the number of sources in our source plane area that have at least one image detected in a $5\arcmin \times 5\arcmin$ area mosaic survey of comparable depth to that of \citet[29.8 mag in WFC3 $H_{160W}$;][]{bouwens2011a} relative to a similar observation in a blank field of equivalent area.  We assume the $z \sim 10$ luminosity function constraints of \citet{bouwens2011a}, who chose fixed values of $\phi^{*} = 0.0012$ Mpc$^{-3}$ and $\alpha = -1.74$ and derived a constraint of $M^{*} = -18.3 \pm 0.5$ mag.

\begin{figure*}
\centering
\plotone{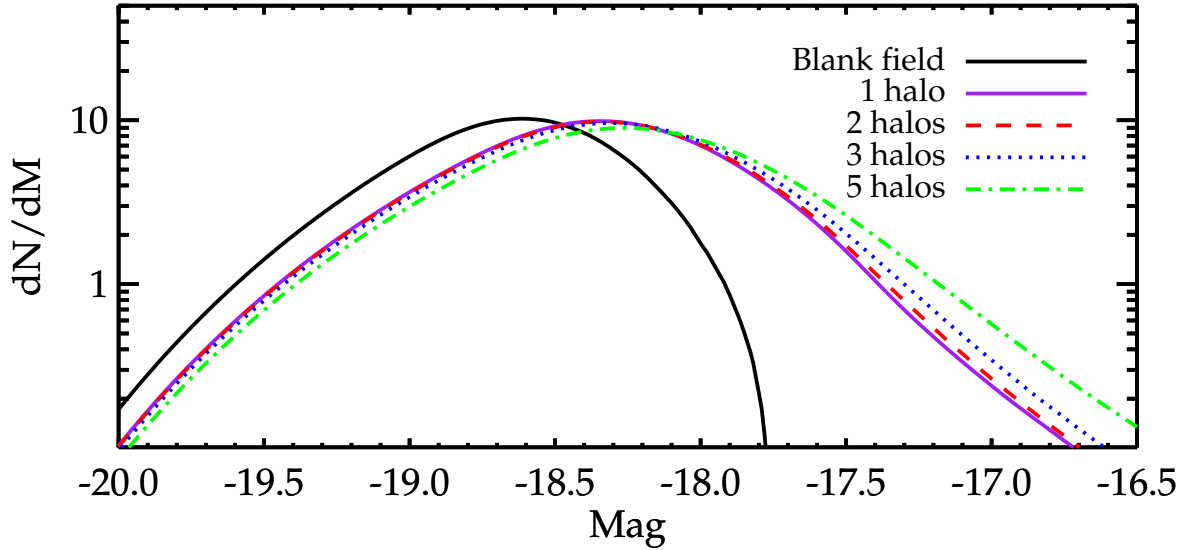}
\caption{Number of $z = 10$ detections per unit magnitude for hypothetical $5\arcmin \times 5\arcmin$ space-based blank field (black lines) and lensing field (colored lines) observations.  The luminosity function of \citet{bouwens2011a} is assumed.  The blank field observation can detect more objects at the bright end due to the larger volume probed in the source plane, while the faint end is only detectable in the lensing fields.  The tradeoff between the bright and faint end tends more toward faint end detections as the number of halos increases, reflecting the gain in the cross section.  The mass configurations with more halos, in addition to having larger cross sections, also probe larger source plane volumes at magnifications needed to detect the faint end of the LF, resulting in more detections of objects fainter than $M^{*}$.  \label{fig:dndm_avg}}
\end{figure*}

In Figure~\ref{fig:dndm_avg}, we plot the number of detections per unit magnitude as a function of source luminosity for both blank fields and our model mass configurations.  The blank field observation can detect more objects at the bright end due to the larger volume probed in the source plane.  The faint end, on the other hand, is beyond its reach, but can be detected in the lensing fields.  The relative tradeoff between these two detected source populations favors faint end detections as the number of halos increases, reflecting the gain in the lensing cross section for these mass configurations.  Therefore, the mass configurations with more halos, in addition to having larger cross sections, also probe larger source plane volumes at the faint end of the LF.  The effects of changing various halo properties are generally smaller than the effect of changing the number of halos.

As a result of the large uncertainty in $M^{*}$ ($\pm 0.5$ mag), the uncertainties in the curves in Figure~\ref{fig:dndm_avg} due to error in the LF are comparable to the differences among the various curves.  For example, in the particular case of $M^{*} = -18.3$, the 1-halo and 5-halo beams result in a factor of $\sim$3.5 and $\sim$4 increase, respectively, in the number of detections fainter than $M^{*}$ over the blank field.  However, changing $M^{*}$ by $1\sigma$ to $-17.8$ results in a factor of $\sim$23 and $\sim$37 increase, respectively, over the number of faint blank field detections.  These large deviations result from the fact that there are very few faint-end detections expected from blank field observations.  Better constraints on both the luminosity and size distributions of high-redshift sources are clearly needed, and can in fact be provided by observations of such multi-halo beams.

We caution against using these numbers as more than rough estimates due to the large LF uncertainties and the strong dependence on the details of the observing program.  For example, ground-based surveys, while still inferior to high-resolution space-based imaging, will benefit more from lensing due to the larger PSFs of current ground-based detectors.  Large uncertainties arise because the assumed depth of the observation only reaches a luminosity that is still in the exponential part of the LF, so a change in $M^{*}$ leads to a large change in the gain due to lensing.  Nevertheless, the improvement due to splitting the beam's mass into more than one halo is clear.

\subsection{Likelihood of Multiple Clusters in Projection} \label{subsec:likelihood}
Although we have demonstrated that lines of sight containing multiple cluster-sized halos nearby in projection can be better at lensing high-redshift galaxies into detectability than those containing a single cluster, it is reasonable to consider the likelihood of such fields existing in the real universe.  A handful of serendipitous discoveries of cluster-cluster lenses {\it have} been reported in the literature \citep{blakeslee2001,athreya2002,zitrin2012}, albeit with smaller total integrated masses than our simulated fields, so their existence is not simply a theoretical construct.  Here, we perform a simple calculation using our assumed halo mass function to estimate the number of fields in the universe that contain two cluster-scale halos with a total integrated mass of at least $2 \times 10^{15} M_{\odot}$.

We begin by performing the calculation of 2-halo beams in general terms, then examine specific cases that are relevant in the context of our simulations.  We focus on the case of two halos for simplicity, requiring that each halo has at least a minimum mass, $M_{min}$, and lies within the redshift range $z_{min} \leq z_{1} \leq z_{2} \leq z_{max}$.  Multi-cluster lines of sight will be strongly biased toward low masses due to the declining mass function, so we also require that the two masses sum to some minimum total mass, $M_{tot}$.  We define the quantity 
\begin{equation} \label{eq:numden}
n(M,z) = \int_{\Omega} \frac{dN}{dM dV} = 4\pi \frac{H(z)}{c} \frac{dN}{dM dz}
\end{equation}
to be the number of halos with mass $M$ at redshift $z$ across the entire sky, where $H(z)$ is the Hubble parameter at redshift $z$.  Using the halo mass function to determine $dN/dM dz$, we then calculate the number density of foreground halos, $n(M_{1},z_{1})$, and the number density of background halos, $n(M_{2},z_{2})$, by evaluating the mass function at the appropriate masses and redshifts.   We account for the maximum separation range of the two halos by requiring that the second halo lies within a circle of angular radius $R_{sep}$.  This is a fixed quantity for all redshifts, and therefore is a constant that factors out of the integral.  If we had chosen a geometry-dependent maximum separation (e.g., the Einstein radius of the foreground lens), we would need to include the solid-angle term when integrating over $z_{2}$.  Our restrictions on both the minimum halo mass and total mass restricts the minimum value of $M_{2}$ to either $M_{min}$ or $(M_{tot} - M_{1})$, whichever is greater.  With these assumptions, the total number of 2-halo lines of sight that have a mass of at least $M_{tot}$ is given by
\begin{equation} \label{eq:likelihood}
\begin{split}
N =& \left( \frac{\pi R_{sep}^{2}}{4 \pi~\mathrm{ster}} \right) \int_{M_{min}}^{\infty} dM_{1} \int_{z_{min}}^{z_{max}} dz_{1} \\
& \int_{\mathrm{max}(M_{min}, M_{tot} - M_{1})}^{\infty} dM_{2} \int_{z_{1}}^{z_{max}} n(M_{1},z_{1}) n(M_{2},z_{2}) dz_{2}.
\end{split}
\end{equation}

With this general calculation, we can then examine specific cases.  We look at the 2-halo case analyzed in our Monte Carlo trials by setting $M_{min} = 10^{14} M_{\odot}$, $M_{tot} = 2 \times 10^{15} M_{\odot}$, $z_{min} = 0.1$, $z_{max} = 0.9$, and $R_{sep} = 3\farcm5$.  Based on this calculation, we expect there to be $\sim$10 such lines of sight in the sky.  This indicates that there exist $\sim$1 field similar to the top 10\% of just the 2-halo beams analyzed in \S~\ref{subsec:top10}, so it is reasonable to expect that such powerful lensing beams do exist.  The frequency of multi-halo lines of sight also increases dramatically if our requirements on total mass or minimum halo mass are relaxed.

The calculation is very sensitive to our assumed cosmology, particularly $\sigma_{8}$.  The uncertainty in $\sigma_{8}$ from the seven-year WMAP results suggests a factor of $\sim2$ uncertainty.  This simplistic calculation is likely conservative, as it does not account for mass due to smaller structures along the line of sight, the possibility of more than two cluster-scale halos in projection, or the effects of large-scale structure that could lead to correlations among mass peaks.  Accounting for these effects could increase the predicted total number of massive multi-halo fields.  A more detailed analysis of the frequency of these high-mass lines of sight using the Millenium Simulation will be presented in French et al. (in preparation).  Additionally, we have identified several high-mass lines of sight containing multiple structures from spectroscopic observations of selected SDSS fields.  An ongoing spectroscopic and photometric observing program has discovered and confirmed at least two lines of sight that contain multiple halos each with dynamical masses greater than $\sim 10^{14} M_{\odot}$ as inferred from the relative velocities of the cluster galaxies.  The total integrated masses in these fields are estimated to be $\gtrsim 3\times10^{15} M_{\odot}$.  The details of these observations will be presented in Ammons et al. (in preparation).

\section{CONCLUSIONS} \label{sec:conclusions}
We have compared the lensing properties of lines of sight containing chance alignments of multiple cluster-scale halos on the sky to single-halo lenses, isolating the effects of more projected halos and the detailed properties of those halos.  We examine the implications for magnifying very high-redshift ($z \sim 10$) galaxies.  Our main conclusions are as follows.

\begin{itemize}
\item Lines of sight containing multiple cluster-scale halos have larger cross-sections for lensing high-redshift galaxies above a given magnification threshold than single halos of equivalent total integrated mass.  This effect is driven by interactions among the lens potentials of the halos, arising from non-linear combinations of both their convergence and shear profiles.  The gain in the lensing strength due to the presence of multiple halos is maximized at some finite projected separation.  For a beam with a total mass of $2\times10^{15} M_{\odot}$, this separation is $\sim$100\arcsec.  At larger separations, the configuration approaches the limit where the halos are independent fields.  At smaller separations, the mass distribution is inefficient at maximizing the field's \'{e}tendue (the area containing large magnifications), resulting in a lower cross section.
\item For a line of sight containing the same total integrated mass split between two cluster-scale halos, the gain in lensing cross section is larger in those rare cases where the halo masses are comparable.
\item The physical properties of a halo that drive an increase in the lensing cross section are a high concentration, the major axis aligned with the line of sight, and an intermediate redshift ($\gtrsim 0.4$), which optimizes the scaled surface density of the halo.  The triaxiality and projected ellipticity of the halo, and the relative orientations of multiple halos, are relatively unimportant.  For beams with two cluster-scale halos, one halo typically dominates the mass due to the shape of the halo mass function.  In such cases, the properties of the dominant halo determine the lensing characteristics of the field.  One implication of this result is that analyses of known single-halo lensing beams, which rarely account for other structures along the line of sight, are unlikely to have a second massive halo in the beam that would significantly alter the overall lensing cross section, but errors in the derived 2-D magnification map may still be underestimated.
\item The same factors that increase the lensing cross section in typical 2-halo beams generally produce the best (top 10\%) of 2-halo beams.  One exception is that halos at low redshifts that scatter to large concentrations, resulting in an increased surface density over a large area, can have unusually large lensing cross sections.  This is opposite to the overall trend for halos at low redshift to have smaller cross sections.
\item We have isolated the effects of mass configuration $-$ multiple halos and the properties of those halos $-$ while controlling for total integrated mass in the beam.  These effects are non-negligible and can be comparable to increasing the total mass in the beam from $\sim10^{15} M_{\odot}$ to $\sim3\times10^{15} M_{\odot}$.  Actual lines of sight in the universe containing multiple halos have the additional benefit that they are likely to contain more total mass than those with just a single massive cluster (French et al. in preparation).
\item We have conservatively estimated the number of lines of sight that contain two cluster-scale halos within $0.1 \leq z \leq 0.9$ as $\sim$10 in the sky.  A more detailed constraint using cosmological simulations will be presented in a future work (French et al. in preparation).  We have identified several fields in the SDSS with large integrated masses and multiple cluster-scale halos in projection.  Detailed observations of these fields will be presented in Ammons et al. (in preparation).
\end{itemize}

The results of our analysis can be used to identify ideal lines of sight in the real universe for magnifying very high-redshift galaxies into detectability once the masses, redshifts, centroids, and certain physical properties (e.g., concentration) of multiple cluster-scale halos are constrained from observational data.  This work shows that focusing on reducing the uncertainties in these parameters will most limit the errors in the resulting magnification maps, which in turn will improve constraints on the luminosity function of lensed sources at $z \sim 10$ identified in these fields.

\acknowledgments
We thank Decker French for her contributions to this project.  We also thank Michael Blanton, Romeel Dav\'{e}, Daniel Eisenstein, Brenda Frye, David Hogg, Daniel Marrone, and Johan Richard for helpful discussions and input, as well as Masamune Oguri for providing information on the projection of triaxial halos.  This work was supported by NSF grant AST-0908280.  This work performed in part under the auspices of the U.S. Department of Energy by Lawrence Livermore National Laboratory under Contract DE-AC52-07NA27344.  A.I.Z thanks the Center for Cosmology and Particle Physics at New York University for their hospitality and support during her stays there.

\appendix
\section{CHOICE OF $\sigma_{\mu}$ MAGNIFICATION THRESHOLD} \label{app:muthreshold}
Our chosen threshold of $\mu_{t} = 3$ for the brightest image of a given source position is somewhat arbitrary, although there is evidence that the number of detections in lensing fields as a function of magnification broadly peaks around $\mu \sim 3$ (Ammons et al. in preparation).  However, one could imagine that for the purposes of magnifying high-redshift galaxies, we may want a different magnification threshold.  At $z \sim 10$, realistic observational programs with $HST$ or a future facility will have little or no ability to probe significantly fainter than $M^{*}$.  In the exponential part of the luminosity function, the gain in source density outweighs the loss of volume probed for increasing lensing magnification, so higher magnification may be desirable.

\begin{figure}
\centering
\plottwo{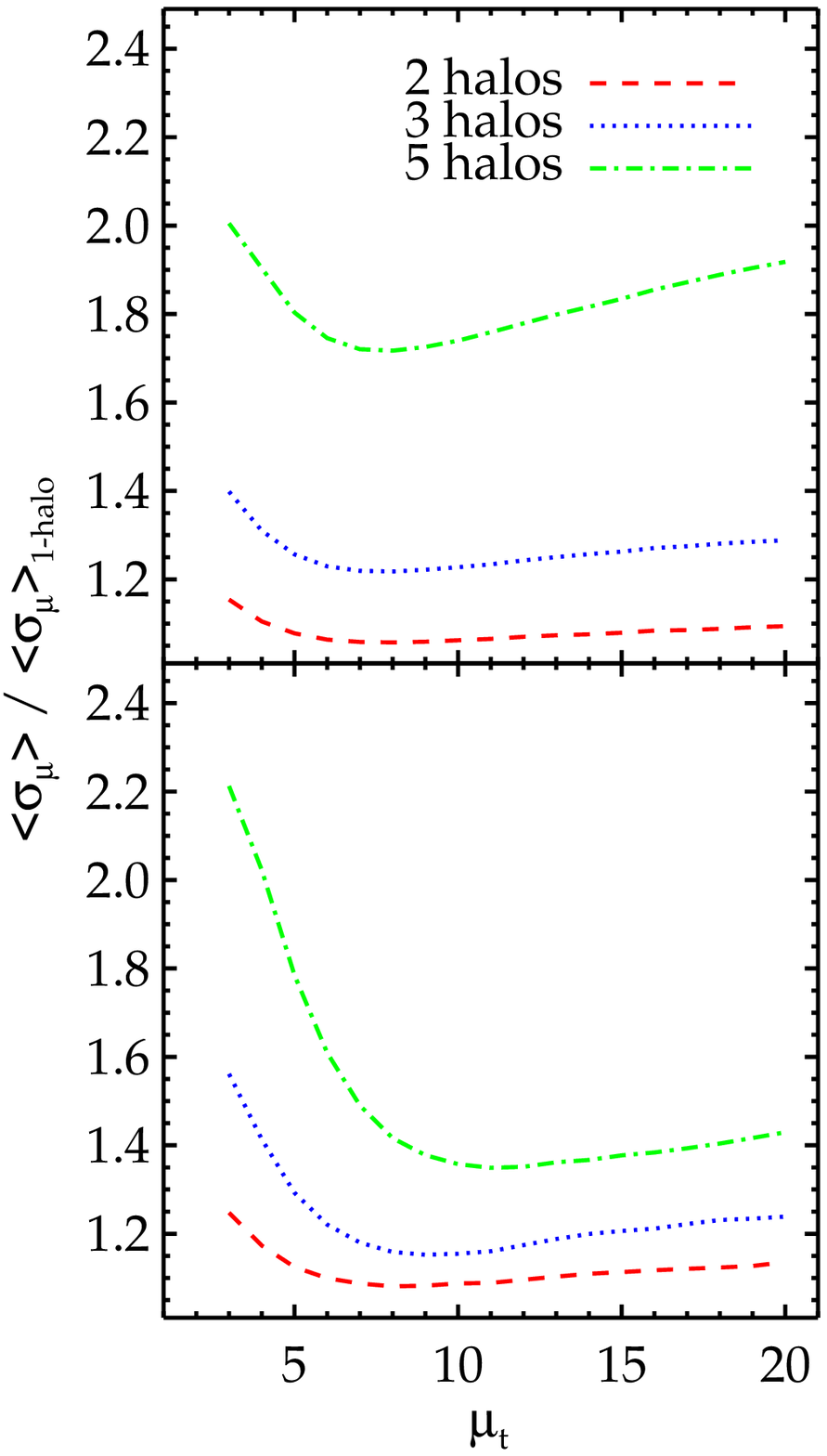}{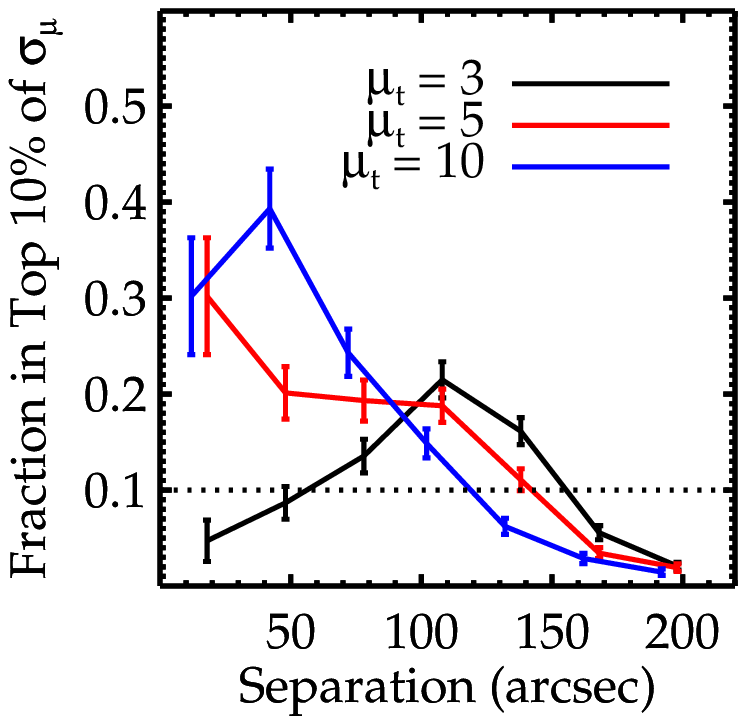}
\caption{{\it Left:} $\sigma_{\mu}$ relative to the 1-halo case as a function of the threshold magnification, $\mu_{t}$ at which $\sigma_{\mu}$ is defined for each of the beam configurations (top).  We make a similar plot, including only the top 10\% of beams ranked by $\sigma_{\mu > \mu_{t}}$ (bottom). The gain in cross section for the multi-halo fields over the single-halo fields is always positive on average, regardless of our choice of magnification threshold for defining $\sigma_{\mu}$ in this range.  {\it Right:} Same as the left panel of Figure~\ref{fig:top10} but for $\sigma_{\mu}$ defined to include magnifications above 3 (black), 5 (red), and 10 (blue).  The peak shifts to smaller separations as $\mu_{t}$ increases, reflecting the fact that larger projected surface densities are needed to produce regions of at least the threshold magnification value. \label{fig:sigmamutrend}}
\end{figure}

We reevaluate $\sigma_{\mu}$ for different magnification thresholds, ranging from $\mu_{t} = 3$ to $\mu_{t} = 20$, and plot the mean values relative to the single-halo case in the top panel of the left plot of Figure~\ref{fig:sigmamutrend}.  We can see that the choice of magnification threshold within this range does not change our general result that the multi-halo fields have larger cross-sections than the single-halo case, although the fractional gain over the single-halo case does vary.  In the bottom panel, we plot the same quantity for just the top 10\% fields.  We do not examine higher magnification thresholds because the chosen resolution of our source plane grid starts to become problematic, and it would require a large amount of computation time to reevaluate these configurations on a more finely sampled grid.  Likewise, smaller magnification thresholds ($1 \leq \mu_{t} \leq 3$) require a calculation over a larger source plane area and are starting to become less scientifically interesting since they are approaching the case of no lensing magnification.  The results appear unlikely to change at low redshift anyway given the nature of the trends.

Another possible consequence of changing $\mu_{t}$ is that the the effects of various parameters on $\sigma_{\mu}$, such as the ones discussed in \S~\ref{subsec:params}, may change.  In general, increasing this threshold will reduce the normalization of $\sigma_{\mu}$, but it is uncertain whether its behavior as a function of these parameters will change.  Performing the simple mass scaling test shown in Figure~\ref{fig:sigmamass} for different choice of $\mu_{t}$ only appears to change the overall normalization, but the nature of the trend remains the same.  What about the other physical parameters that were explicitly tested in \S~\ref{subsec:params}?

We test the effects of changing $\mu_{t}$ for various physical parameters and find that they are qualitatively similar with the exception of angular separation.  In the right panel of Figure~\ref{fig:sigmamutrend}, we replot the left panel of Figure~\ref{fig:top10}, but with the results for $\mu_{t} = 5$ and $\mu_{t} = 10$ as well.  We plot this figure rather than one analogous to the upper left panel of Figure~\ref{fig:sigmamusepmr} to remove the effect of overall normalization, and because the trends with separation that applies to the top 10\% of beams are qualitatively similar to the mean results.  As we increase $\mu_{t}$, the optimal angular separation decreases.  This makes sense because in order for interactions between the lens potentials to produce extended magnifications regions above an increasing threshold, the interactions between the convergence and shear terms of the different halos need to be stronger.  This indicates that the halos need to be projected closer together since their projected density and shear profiles drop off with increasing radius.

\section{PROJECTION EFFECTS ON HALO PROPERTIES} \label{app:projection}
The lensing properties of a given mass distribution are determined by its projected mass density in the plane of the sky.  In our analysis, we assume NFW halos with a triaxial mass distribution, which we need to translate into projected quantities.  Here, we describe our method for characterizing the projected mass distribution of a triaxial ellipsoid \citep[see also][]{dezeeuw1989,oguri2003,vandeven2009,morandi2011,feroz2012}.

For a spherical NFW halo, the density profile is given by 
\begin{equation} \label{eq:nfw_profile}
\rho(r) = \frac{\rho_{s}}{(r/r_{s})(1+r/r_{s})^{2}}
\end{equation}
where $r_{s}$ is the scale radius where the slope of the mass profile matches that of an isothermal sphere, and $\rho_{s}$ is the characteristic density.  When projected onto the plane of the sky, the surface density profile is given by 
\begin{equation} \label{eq:nfw_surf}
\begin{split}
  \Sigma_{0}(R) &= \int_{-\infty}^{\infty} \rho\left(\sqrt{R^{2}+z^{2}}\right)\,dz \\
  &= 2 \int_{0}^{\infty} \rho(R\cosh u)\,R\cosh u\,du
\end{split}
\end{equation}
with the change of variables $z = R\sinh u$.  Since $\rho(r)$ depends on its argument only through $r/r_{s}$, we know that $\Sigma_{0}(R)$ depends on $R$ only through the combination $R/r_{s}$.

For a triaxial NFW halo, we specify its major, intermediate, and minor semi-axes by $a$, $b$, and $c$, respectively.  To simplify the notation in this section, we rescale these semi-axis ratios by $r_{s}$ such that $a \equiv a/r_{s}$, $b \equiv b/r_{s}$, and $c \equiv c/r_{s}$ are dimensionless quantities.  Let $\mbox{\boldmath $X$} = (X, Y, Z)$ be coordinates in the halo's principal axis frame, and $\mbox{\boldmath $x$}=(x, y, z)$ be coordinates in the observer's frame such that the $z$-axis is along the line of sight.  The two coordinate systems are related by a rotation matrix,
\begin{equation} \label{eq:rot}
  \mbox{\boldmath $X$} = {\bf A} \mbox{\boldmath $x$}.
\end{equation}
We then define
\begin{equation} \label{eq:msq}
  m^2 = \frac{X^2}{a^2} + \frac{Y^2}{b^2} + \frac{Z^2}{c^2},
\end{equation}
where $m$ has dimensions of length and is analogous to $r$ in the spherical case.  We can define a triaxial NFW halo by substituting $m$ into Equation~\ref{eq:nfw_profile} to get $\rho(m)$.  Using Equation~\ref{eq:rot}, we can write $m^{2}$ as a quadratic in $(x, y, z)$:
\begin{equation} \label{eq:m}
m^{2} = f z^{2} + g z + h, 
\end{equation}
where $f$, $g$, and $h$ are functions of $x$, $y$, and the angles used to define the halo orientation.  Explicit expressions for these quantities are given in \S~\ref{subsec:viewing} and~\ref{subsec:euler}.  We rewrite Equation~\ref{eq:m} as
\begin{equation} \label{eq:m2}
m^{2} = \xi^{2} + \zeta^{2},
\end{equation}
where
\begin{equation} \label{eq:xi}
\xi = f^{1/2} \left( z + \frac{g}{2f} \right),
\end{equation}
and
\begin{equation} \label{eq:zeta}
\zeta = \sqrt{h-\frac{g^{2}}{4f}}.
\end{equation}
Now when we do the projection, we can change integration variables from $z$ to $\xi$.  All of the $(x, y)$ dependence is in $\zeta$, and $\zeta^{2}$ is quadratic in $(x, y)$ such that there is some rotated coordinate system $(x^{\prime}, y^{\prime})$ where
\begin{equation} \label{eq:zetasq_rot}
\zeta^{2} = \left(\frac{x^{\prime}}{a^{\prime}} \right)^{2} + \left( \frac{y^{\prime}}{b^{\prime}} \right)^{2}.
\end{equation}
This means that any function of $\zeta$ will have elliptical symmetry.  The quantities $a^{\prime}$ and $b^{\prime}$ are the semi-major and semi-minor axes (again scaled by $r_{s}$ so they are dimensionless quantities), respectively, of the projected ellipse.  The frames are rotated by some angle $\psi$, so
\begin{equation} \label{eq:rot_mat}
  \left[\begin{array}{c} x' \\ y' \end{array}\right] =
  \left[\begin{array}{rr} \cos\psi & \sin\psi \\ -\sin\psi & \cos\psi \end{array}\right]
  \left[\begin{array}{c} x \\ y \end{array}\right].
\end{equation}
We can rearrange Equation~\ref{eq:zeta} and write it as
\begin{equation} \label{eq:zetasq2}
\zeta^{2} = d_{xx} x^{2} + d_{xy} x y + d_{yy} y^{2},
\end{equation}
where $d_{xx}$, $d_{xy}$, and $d_{yy}$ are the coefficients of the $x^{2}$, $xy$, and $y^{2}$ terms, respectively.  Equating Equation~\ref{eq:zetasq2} with Equation~\ref{eq:zetasq_rot} and substituting for $x^{\prime}$ and $y^{\prime}$ via Equation~\ref{eq:rot_mat}, we can solve for $\psi$, $a^{\prime}$, and $b^{\prime}$:
\begin{align} \label{eq:abpsi}
  \psi &= \frac{1}{2} \arctan \left( {\frac{d_{xy}}{d_{xx}-d_{yy}}} \right), \\
  a' &= \sqrt{\frac{2}{F-G}}, \\
  b' &= \sqrt{\frac{2}{F+G}},
\end{align}
where
\begin{align} \label{eq:fg}
F &= d_{xx}+d_{yy}, \\
  G &= \sqrt{(d_{xx}-d_{yy})^2+d_{xy}^2}.
\end{align}

We can implement one of two equivalent approaches using this basic framework.  We can specify the orientation of the halo in terms of two viewing angles and apply a random rotation on the sky at the end.  Alternatively, we can specify the halo orientation in terms of three Euler angles which will naturally produce a random projected orientation.

\subsection{Viewing Angle Approach} \label{subsec:viewing}
Three angles are needed to fully specify the orientation of a halo in three dimensions.  In the viewing angle approach, we only specify two angles, so the projected mass distribution must be rotated by a random angle at the end to achieve different position angles.  In this approach, we define $\theta$ and $\phi$ to be the polar and azimuthal angles, respectively, of the ellipsoid relative to the observer's frame.  Following the formalism in \citet{oguri2003}, the rotation matrix {\bf A} is given by
{\setlength{\arraycolsep}{10pt}
\begin{equation} \label{eq:amat}
\begin{split}
{\bf A} &= \left[
\begin{array}{ccc}
 -\sin \phi & -\cos \theta \cos \phi & \sin \theta \cos \phi \\
 \cos \phi & -\cos \theta \sin \phi & \sin \theta \sin \phi \\
 0 & \sin \theta & \cos \theta
\end{array}
\right].
\end{split}
\end{equation}
}
The coefficients in Equation~\ref{eq:m} are given by 
\begin{align} \label{eq:fgh}
  f &= \sin ^2 \theta \left(\frac{\cos^{2} \phi}{a^{2}}+\frac{\sin^{2} \phi}{b^{2}}\right)+\frac{\cos^{2} \theta}{c^{2}}, \\
  g &= \sin \theta \sin 2\phi \left(\frac{1}{b^{2}}-\frac{1}{a^{2}}\right) x + \sin 2\theta \left(\frac{1}{c^{2}}-\frac{\cos^{2} \phi}{a^{2}}-\frac{\sin^{2} \phi}{b^{2}}\right) y, \\
  h &= \left( \frac{\sin^{2} \phi}{a^{2}}+\frac{\cos^{2} \phi}{b^{2}} \right) x^{2} + \left[ \cos^{2} \theta \left(\frac{\cos^{2} \phi}{a^{2}}+\frac{\sin^{2} \phi}{b^{2}}\right)+\frac{\sin^{2} \theta}{c^{2}}\right] y^{2} + \cos \theta \sin 2\phi \left(\frac{1}{a^{2}}-\frac{1}{b^{2}}\right) x y.
\end{align}
Proceeding with these definitions, the coefficients in Equation~\ref{eq:zetasq2} are given by
\begin{align} \label{eq:dxy}
  d_{xx} &= \frac{1}{D} \left[ a^{2} \cos^{2} \theta \cos^{2} \phi + b^{2} \cos^{2} \theta \sin^{2} \phi + c^{2} \sin^{2} \theta \right], \\
  d_{yy} &= \frac{1}{D} \left[ a^{2} \sin^{2} \phi + b^{2} \cos^{2} \phi \right], \\
  d_{xy} &= \frac{1}{D} (b^{2}-a^{2}) \cos \theta \sin 2\phi, \\
  D &= a^{2} b^{2} \cos^{2} \theta + c^{2} \sin^{2} \theta (a^{2} \sin^{2} \phi + b^{2} \cos^{2} \phi).
\end{align}
The angle $\psi$ is not relevant in this approach because we have only specified two angles, which is not sufficiently general to fully characterize the halo orientation.  Once the axis ratios of the projected ellipse have been determined, the ellipse can be rotated by a random angle to achieve different position angles.

\subsection{Euler Angle Approach} \label{subsec:euler}
Alternatively, we can specify the orientation of the halo by three Euler angles, $\alpha$, $\beta$, and $\gamma$.  These angles are related to the viewing angles $\theta$ and $\phi$ by the transformation
\begin{align} \label{eq:angle_trans}
  \alpha &= 0, \\
  \gamma &= -\phi - \frac{\pi}{2}, \\
  \beta &= -\theta.
\end{align}
In this formulation, the rotation matrix becomes
{\setlength{\arraycolsep}{10pt}
\begin{equation} \label{eq:amat_euler}
\begin{split}
  {\bf A} &= \left[
\begin{array}{ccc}
 \cos \alpha \cos \gamma-\sin \alpha \cos \beta \sin \gamma & \cos \alpha \cos \beta \sin \gamma+\sin \alpha \cos \gamma & \sin \beta \sin \gamma \\
 -\sin \alpha \cos \beta \cos \gamma-\cos \alpha \sin \gamma & \cos \alpha \cos \beta \cos \gamma-\sin \alpha \sin \gamma & \sin \beta \cos \gamma \\
 \sin \alpha \sin \beta & -\cos \alpha \sin \beta & \cos \beta
\end{array}
\right].
\end{split}
\end{equation}
}
The quantities $f$, $g$, and $h$ can then be expressed in terms of the three Euler angles as
\begin{align} \label{eq:fgh_euler}
  f =& \sin^{2}\beta \left(\frac{\sin^{2}\gamma}{a^{2}}+\frac{\cos^{2}\gamma}{b^{2}}\right)+\frac{\cos^{2}\beta}{c^{2}}, \\
  g =& \left[\left(\frac{1}{a^{2}}-\frac{1}{b^{2}}\right) \cos \alpha \sin \beta \sin 2\gamma -\sin \alpha \sin 2\beta \left(\frac{\sin^{2}\gamma}{a^{2}}+\frac{\cos^{2}\gamma}{b^{2}}-\frac{1}{c^{2}}\right)\right] x \\
  & + \left[\left(\frac{1}{a^{2}}-\frac{1}{b^{2}}\right) \sin \alpha \sin \beta \sin 2\gamma +\cos \alpha \sin 2\beta \left(\frac{\sin^{2}\gamma}{a^{2}}+\frac{\cos^{2}\gamma}{b^{2}}-\frac{1}{c^{2}}\right)\right] y, \nonumber \\
  h =& \left[ \frac{(\cos \alpha \cos \gamma-\sin \alpha \cos \beta \sin \gamma)^{2}}{a^{2}}+\frac{(\sin \alpha \cos \beta \cos \gamma+\cos \alpha \sin \gamma)^{2}}{b^{2}}+\frac{\sin^{2}\alpha \sin^{2}\beta}{c^{2}}\right] x^{2} \\
  & + \left[ \frac{(\cos \alpha \cos \beta \sin \gamma+\sin \alpha \cos \gamma)^{2}}{a^{2}}+\frac{(\cos \alpha \cos \beta \cos \gamma-\sin \alpha \sin \gamma)^{2}}{b^{2}}+\frac{\cos^{2}\alpha \sin^{2}\beta}{c^{2}}\right] y^{2} \nonumber\\
  & + \left[ \frac{2 (\cos \alpha \cos \beta \sin \gamma+\sin \alpha \cos \gamma) (\cos \alpha \cos \gamma-\sin \alpha \cos \beta \sin \gamma)}{a^{2}} \right. \nonumber\\
  & \quad \left. +\frac{2 (\sin \alpha \cos \beta \cos \gamma+\cos \alpha \sin \gamma) (\sin \alpha \sin \gamma-\cos \alpha \cos \beta \cos \gamma)}{b^{2}}-\frac{\sin 2\alpha \sin^{2}\beta}{c^{2}} \right] x y. \nonumber
\end{align}
The coefficients $d_{xx}$, $d_{xy}$, and $d_{yy}$ in this approach are 
\begin{align} \label{eq:dxy_euler}
  d_{xx} &= \frac{1}{D} \left[ (\sin \alpha \cos \gamma + \cos \alpha \cos \beta \sin \gamma)^{2} a^{2}
    + (\sin \alpha \sin \gamma - \cos \alpha \cos \beta \cos \gamma)^{2} b^{2}
    + (\cos \alpha \sin \beta)^{2} c^{2} \right], \\
  d_{yy} &= \frac{1}{D} \left[ (\cos \alpha \cos \gamma - \sin \alpha \cos \beta \sin \gamma)^{2} a^{2}
    + (\cos \alpha \sin \gamma + \sin \alpha \cos \beta \cos \gamma)^{2} b^{2}
    + (\sin \alpha \sin \beta)^{2} c^{2} \right], \\
  d_{xy} &= \frac{1}{D} \left[ \sin 2\alpha \sin^{2} \beta \left(c^{2}-\frac{a^{2}+b^{2}}{2} \right) 
    - (a^{2}-b^{2}) \left( \frac{1}{4}(3+\cos 2\beta) \sin 2\alpha \cos 2\gamma
      + \cos 2\alpha \cos \beta \sin 2\gamma \right) \right], \\
  D &= a^{2} b^{2} \cos^{2} \beta + c^{2} \sin^{2} \beta (a^{2} \cos^{2}\gamma + b^{2} \sin^{2} \gamma).
\end{align}
The angle $\psi$ (Equation~\ref{eq:abpsi}) is now the position angle of the projected surface mass distribution, whereas a random position angle was drawn independently in the viewing angle approach.  Both approaches give equivalent distributions of projected ellipticities and position angles.

\subsection{Effects on the Density Profile} \label{subsec:effects}
The relevant physical quantities for determining the lensing properties of an NFW halo are the ellipticity $\epsilon$, the scale radius $r_{s}$, and the scaled surface density $\kappa_{s} = r_{s}\rho_{s} / \Sigma_{c}$, where $\Sigma_{c}$ is the critical surface density for lensing.  Assuming a triaxial rather than spherical halo does not change its virial radius or mass, but does change $\epsilon$ and the projected $r_{s}$ and $\kappa_{s}$.  The projected ellipticity is simply
\begin{equation} \label{eq:ellipticity}
\epsilon = 1 - \frac{b^{\prime}}{a^{\prime}}.
\end{equation}
To obtain the surface density profile of a triaxial halo, we substitute $\zeta$ into Equation~\ref{eq:nfw_surf}:
\begin{equation} \label{eq:surf_zeta}
\begin{split}
  \Sigma(\zeta) &= \int_{-\infty}^{\infty} \rho(m)\,dz \\
  &= f^{-1/2} \int_{-\infty}^{\infty} \rho\left(\sqrt{\xi^2+\zeta^2}\right)\,d\xi \\
  &= f^{-1/2} \times 2 \int_{0}^{\infty} \rho(\zeta\cosh u)\,\zeta\cosh u\,du \\
  &= f^{-1/2} \times \Sigma_0(\zeta),
\end{split}
\end{equation}
with the change of variables $\xi = \zeta \sinh u$.  In other words, we can obtain the function for the surface mass density of the elliptical case just by plugging the elliptical radius $\zeta$ into the function $\Sigma_{0}$ and rescaling by $f^{-1/2}$.  We know that the position coordinates enter only through the combination $\zeta/r_{s}$.  In the coordinates aligned with the major axis of the ellipse, this has the form
\begin{equation} \label{eq:zeta_scaled}
  \frac{\zeta}{r_{s}} = \left[ \left(\frac{x'}{a' r_{s}}\right)^2 + \left(\frac{y'}{b' r_{s}}\right)^2 \right]^{1/2}.
\end{equation}
We can therefore think of the projected mass distribution having some scale radius $r_{s}^{\prime}$, and if we choose this so a circle of radius $r_{s}^{\prime}$ has the same area as the ellipse above, then we have
\begin{equation} \label{eq:r_s_prime}
r_{s}^{\prime} = r_{s} \sqrt{a^{\prime} b^{\prime}}.
\end{equation}
Since the virial radius of the halo remains the same, the projected concentration becomes
\begin{equation} \label{eq:c_prime}
c^{\prime} = \frac{r_{vir}}{r_{s}^{\prime}} = \frac{c_{vir}}{\sqrt{a^{\prime} b^{\prime}}}.
\end{equation}
The scaled surface density $\kappa_{s}$ is rescaled by a factor of $f^{-1/2}$  as per Equation~\ref{eq:surf_zeta}, and can also be expressed as a combination of 3D and projected axis lengths \citep{vandeven2009},
\begin{equation} \label{eq:kappa_s_prime}
\kappa_{s}^{\prime} = f^{-1/2} \kappa_{s} = \frac{abc}{a^{\prime} b^{\prime}} \kappa_{s}.
\end{equation}

\bibliography{toymodel}

\end{document}